\documentclass[aps,pra,showpacs,twocolumn,amsmath,amssymb,superscriptaddress,footinbib]{revtex4}

\usepackage[english]{babel}
\usepackage{ifthen}
\usepackage{latexsym}
\usepackage{graphicx}
\usepackage{epsfig}
\usepackage[T1]{fontenc}
\usepackage[usenames,dvipsnames]{color}
\usepackage{ulem}

\def\be{\begin{equation}}
\def\ee{\end{equation}}
\def\bea{\begin{eqnarray}}
\def\eea{\end{eqnarray}}
\def\bi{\begin{itemize}}
\def\ei{\end{itemize}}
\def\bin{\begin{enumerate}}
\def\ein{\end{enumerate}}

\begin{document}

\title{Numerical Computation of Dynamically Important Excited States of Many-Body Systems}

%%%%%%%%%%%%%%%%%%%%%%%%%%%%%%%%%%%%%%%%%%%%%%%%%%%%%%%%%%%%%%%%%%%%%%%%%%%%%%%
\author{Mateusz \L\k{a}cki}

\affiliation{
Instytut Fizyki imienia Mariana Smoluchowskiego, 
Uniwersytet Jagiello\'nski, ulica Reymonta 4, PL-30-059 Krak\'ow, Poland}

\author{Dominique Delande}

\affiliation{Laboratoire Kastler-Brossel, UPMC-Paris 6, ENS, CNRS;  
4 Place Jussieu, F-75005 Paris, France}
 
\author{Jakub Zakrzewski} 

\affiliation{
Instytut Fizyki imienia Mariana Smoluchowskiego, 
Uniwersytet Jagiello\'nski, ulica Reymonta 4, PL-30-059 Krak\'ow, Poland}

\affiliation{
Mark Kac Complex Systems Research Center, 
Uniwersytet Jagiello\'nski, Krak\'ow, Poland}

\date{\today}

\begin{abstract}

We present an extension of the time-dependent Density Matrix Renormalization Group (t-DMRG), 
also known as Time Evolving Block Decimation algorithm (TEBD), allowing for the
computation of dynamically important excited states of one-dimensional many-body systems. 
We show its practical use for analyzing the dynamical properties and excitations
of the Bose-Hubbard model describing ultracold atoms loaded 
in an optical lattice from a~Bose-Einstein condensate.
This allows for a~deeper understanding of nonadiabaticity in experimental realizations of insulating phases.  
\end{abstract}


\maketitle
%%%%%%%%%%%%%%%%%%%%%%%%%%%%%%%%%%%%%%%%%%%%%%%%%%%%%%%%%%%%%%%%%%%%%%%%%%%%%%%

\section{Introduction}

Properties of many-body quantum systems are difficult to compute, because the eigenstates
are usually unknown and the large size of the Hilbert space makes even approximate descriptions
difficult to manipulate. Some numerical
methods - such 
as variational methods or mean-field theories - are intrinsically approximate
yielding  
simple expressions
describing most of the interesting physical properties, especially in the thermodynamic limit.
The so-called ``quasi-exact'' methods do not have such restrictions, and are supposed
to converge to the exact quantum result when some parameter tends to infinity. 
Those include exact diagonalization methods and 
 quantum Monte-Carlo methods \cite{cep86,bauer11}  (the
limiting
parameter being the statistical sampling). 
For one-dimensional (1D) systems, particularly effective is the Density Matrix Renormalization Group
(DMRG) approach~\cite{white92,hallberg06}. The states, in modern implementations of the method,
are represented by the so called Matrix Product States (MPS) (for a recent smooth introduction, 
see \cite{scholl}). 
In numerous quantum systems, the ground state is very well approximated
by a~MPS, allowing its efficient description with a~relatively small number of parameters, 
which, moreover, does
not increase exponentially with the system size. To some extent, 
DMRG shares the advantages of both variational and quasi-exact methods.

While the original application of these methods considered stationary properties, recent years brought a big progress in addressing
the dynamics. 
Arguably, it was the formulation of the Time Evolving Block Decimation (TEBD) algorithm \cite{vidal03}, 
enabling study of real time dynamics of 1D systems of a reasonable size, which triggered a rapid 
progress in the field. Soon, a time-dependent DMRG
(t-DMRG) method was proposed \cite{daley,white04} with a fast realization of a close equivalence 
between TEBD and t-DMRG.
Various applications of the method addressed dynamics of quantum phase transitions 
\cite{clark04,cuciap,zakrzewski09}, quantum quenches and 
thermalizations~\cite{anna,bernier11,biroli11,bernier12}, 
periodic driving \cite{kollath06,huo11}, to give just a few examples.

As soon as the evolution of the system ceases to be adiabatic, 
the excited states become important in the dynamics. The global
properties of whole groups of excited states may be obtained  
from evaluation of e.g. the spectral function. 
Following the pioneering work of Hallberg
\cite{hallberg95}, a lot of effort has been put in developing efficient schemes for spectral functions 
evaluations (see e.g. \cite{white08,pereira09,barthel09,dargel11,holzner11}
and references therein).  

In the present paper, our aims are somehow complementary. We show how one may obtain selected, 
dynamically important excited eigenstates of strongly correlated systems in a relatively simple way. 
Instead of addressing spin chain systems, as typical for tests of various DMRG improvements,
we shall immediately turn our attention to cold atoms settings and consider a Bose-Hubbard model 
as realized by cold atoms in 1D optical lattice potentials.  
In~\cite{zakrzewski09}, we have shown on such a specific example 
how to extract the energies of the few lowest excited states of the system,
with, however, no access to their properties. In the present paper, 
we explain  how to use  the TEBD (or t-DMRG)  method in order
to extract  some of the excited states with quite good accuracy. 
Those states are expressed also  in terms of MPS  making the calculations
of expectations values and correlation functions easy. 
While the area law theorem \cite{area} assures that the ground state may be efficiently represented 
in MPS representation this is by no means guaranteed for a generic excited state. 
Our method will work only for those states for which such an efficient MPS representation exists.
We shall show, however, that this restriction leaves plenty of interesting ground to explore as,
most often, many excited states up to a reasonably large excitation energy, are also efficiently
expressed as MPSs.
Interestingly,  the method selects states that are
important in the systems' dynamics. It seems more suited for low lying excited states 
(although we give other examples too). Still, in many-body systems, the density of states 
increases usually very fast with energy, and even relatively small systems 
(such as e.g. encountered in ultracold atomic gases) have plenty of excited
states lying not far above the ground state energy.

While we use as an example 
the Bose-Hubbard (BH) model describing an ultra-cold atomic bosonic gas
loaded in a~one-dimensional optical lattice, the
 method proposed  is quite general and can be used for any system where TEBD (t-DMRG) works for 
sufficiently long times. The latter
restriction is necessary since the method is based on Fourier Transform techniques with resolution 
being dependent on the integration time.
We shall show later that the method has an unexpected self purifying property allowing sometimes 
to use much longer integration times than anticipated in standard applications of the 
algorithm \cite{vidal03,daley}. Possible future improvements of the method are discussed 
in the concluding section. 

In the following sections, we first describe the method in some details. Then, we apply 
it for extracting selected excited 
states populated either by the periodic driving or by the non adiabatic dynamics in 
the disordered Bose-Hubbard system. The illustrative examples of excited states show
the power and the limitations of our approach.

\section{Extraction of eigenstates by Fourier Transform} 
 
The basic idea is to start from an initial state which is a~``wavepacket'', 
that is a~superposition of various eigenstates, described by a~MPS, 
to propagate it in real time with a~time-independent
Hamiltonian, finally to combine the MPS states at various times to reconstruct the excited states. 
The evolution of a~state by a~time-independent Hamiltonian with eigenbasis {$|e_i\rangle$} is given by
$|\psi(t)\rangle = \sum_i \exp(-i E_i t/\hbar) c_i |e_i\rangle.$
where $c_i=\langle e_i|\psi(0)\rangle,$
The Fourier transform (FT) of the autocorrelation
function $C(t)=\langle \psi(0)|\psi(t)\rangle = \sum_i |c_i|^2 \exp(-i E_i t/\hbar)$
over a~time interval $T$ is the autocorrelation spectrum:
\begin{equation}
\tilde{C}_T(E) = \frac{1}{T}\!\int_{-T/2}^{T/2}{\!\!\!\!{\mathrm e}^{\frac{i E t}{\hbar}} C(t) dt} = \sum_i{|c_i|^2\
\mathrm{sinc}\frac{(E-E_i)T}{2\hbar}}
\label{eq:ce}
\end{equation}
where $\mathrm{sinc}\,x=\sin x/x.$ In the limit of long time $T$, it
yields narrow peaks at the $E_i$'s  with weights $|c_i|^2,$  that allows to determine the
energy spectrum as shown in~\cite{zakrzewski09}. 

Here we show that  it is actually possible to extract from the dynamics
also the excited eigenstates with large overlap - those that
contribute most to the dynamical wavepacket.
We perform a~FT {\it directly} on the many-body wavefunction $|\psi(t)\rangle:$
\begin{equation}
 |\phi_T(E)\rangle = \frac{1}{T}\! \int_{-T/2}^{T/2}{\!\!\!\!{\mathrm e}^{\frac{i E t}{\hbar}}|\psi(t)\rangle dt}
                   = \sum_i c_i\ \mathrm{sinc}\frac{(E-E_i)T}{2\hbar} |e_i\rangle
\label{eq:phie}
\end{equation}
For long $T, |\phi_T(E_i)\rangle \to c_i|e_i\rangle,$ providing us with the targeted eigenstate.
The
method selects excited states \emph{relevant for the dynamics}, 
although myriads of other states may be present in the same spectral region. 
This paves the way to study the properties of individual excited many-body states.

To evaluate $|\phi_T(E_i)\rangle,$ the integral is approximated by a~discrete series. 
To sum the series, a~procedure to perform the
sum of two MPSs as a MPS is required, which we
now briefly sketch. 
Any state $|\psi\rangle$ of a 1D system of M sites may be expressed in the  MPS 
representation \cite{vidal03,verstraete}:
\begin{equation}
|\psi\rangle\! = \!\!\!\!\!\!\sum\limits_{\genfrac{}{}{0pt}{}{\alpha_1,\ldots,\alpha_M}{i_1,\ldots,i_M}}\!\!\!\!\!
\Gamma_{1\alpha_1}^{[1]
,i_1}\lambda^{[1]}_{\alpha_1}\Gamma^{[2],i_2}_{\alpha_1\alpha_2}\ldots\Gamma^{[M],i_M}_{\alpha_{n-1}1} |i_1,\dots,i_M\rangle
=:\mathcal{M}(\Gamma,\lambda)
\label{eqn:MPS}
\end{equation}
where $\Gamma^{[l],i_l}$ are matrices and $\lambda^{[l]}$ vectors and 
$|i_j\rangle$ span a local Hilbert space on site $j$. 
When the $\Gamma$'s and the $\lambda$'s satisfy simple
orthogonality relations detailed in~\cite{shividal}, their
physical interpretation is simple. For example, the spectrum of the reduced density
matrix associated with bipartite splitting 
$L:R = \{1,\ldots,l\}:\{l+1,\ldots,M\}$, $\text{Tr}_R |\psi \rangle \langle \psi|$ 
is nothing but $\left(\lambda^{[l]}_{\alpha_l}\right)^2$
for $\alpha_l=1,2\ldots.$   The corresponding entanglement entropy given by
\begin{equation}
S_l = - \sum_{\alpha}{ (\lambda^{[l]}_{\alpha})^2 \ln(\lambda^{[l]}_{\alpha})^2 },
\label{eq:ent_ent}
\end{equation}
is an  important tool to characterize the strongly correlated systems,  
e.g. topological phases~\cite{kitaev06}. The collection of $\lambda$ values forms the entanglement spectrum.

To describe exactly a generic state in terms of a MPS, 
a~large number (exponentially increasing with $M$) of $\alpha_l$ values is needed. 
However, quite often,
physical states created in many experiments are only slightly entangled so  
that $\lambda^{[l]}_{\alpha_l=1,2\ldots}$ are rapidly
decaying numbers, which allows for introduction of a~rather small cutoff $\chi$ in all sums above,
resulting in tractable numerical computation \cite{vidal03}. Our approach is limited to such states only.

Following~\cite{mcculloch} one has that:
$\mathcal{M}(\Gamma,\lambda)+\mathcal{M}(\Gamma',\lambda')=\mathcal{M}(\Gamma\oplus\Gamma',\lambda\oplus\lambda').$ 
The result does
not satisfy orthogonality relations that are vital for achieving efficient MPS algorithms. 
The algorithm to restore them~\cite{shividal} ensures  that the sum is stored in a~memory-efficient way. 
Applying the above two-step algorithm to perform elementary addition suffices 
to obtain $|\psi_T(E_i)\rangle.$ The technical details are presented in the Appendix.

Our goal  is to obtain an eigenstate as pure as possible, 
not merely to detect its presence. 
This requires long time integration which may affect the accuracy of the TEBD algorithm \cite{vidal03,daley}. 
The integration time may be significantly shortened using few simple ideas which we now describe.

For simplicity let us restrict to a subspace spanned by two eigenstates 
$|\psi\rangle= \alpha |e_0\rangle + \beta |e_1\rangle$ 
as a special case. 
Performing the Fourier Transform as in eq.~(\ref{eq:phie}) gives:
\begin{equation}
|\psi_T(E_0)\rangle = \alpha |e_0\rangle +  \beta(T) |e_1\rangle, 
\label{eq:psie2}
\end{equation}
 with $\beta(T)=\beta\ \mathrm{sinc}\frac{T\Delta E}{2 \hbar}, \Delta E=E_1 -E_0.$ 
Clearly $\beta(+\infty)=0,$ therefore large enough
$T$ guarantees convergence of $|\psi_T(E_0)\rangle $ to  $|e_0\rangle$.  
Another possibility is to choose $T$ such that
$\beta(T)=0,$ i.e. a multiple of $\frac{4\pi \hbar}{\Delta}.$ 
This ensures that $|\psi_T(E_0)\rangle=|e_0\rangle$ despite relatively
short evolution time. When other eigenstates contribute significantly to the wavepacket, 
there is no possibility to choose such a time
$T$ that contributions from other eigenstates vanish simultaneously. 
However, if for some state $|\psi \rangle,$ and its eigenstate
$|e_{i_0}\rangle$ there exists an eigenstate $|e_{i_1}\rangle$ with nonzero overlap 
on $|\psi\rangle,$ such that $|E_{i_0}- E_{i_1}|
\ll |E_{i_0}-E_j|$ if $j\neq i_0,i_1,$ 
then choosing $T$ as a multiple of $\frac{4\pi \hbar}{\Delta E}$ is an optimal choice. 

If  $\beta\gg \alpha$ in eq.~(\ref{eq:psie2}), a realization of  $\beta(T) \ll 1$ is more difficult. 
One solution is to restart the
Fourier Transform evaluation from a partially converged result. 
Let us first define a map $f_{E,T}(\psi)=\psi_T(E).$ One may
verify that for a two-level system, the $n$ fold composition of $f_{E,T}$ satisfies:

$$f^n_{E,T}(\psi) = f(f^{n-1}_{E,T}(\psi))= \alpha |e_0\rangle + \beta \left(\mathrm{sinc}\frac{\Delta E T}{2\hbar}\right)^n |e_1\rangle $$
This requires performing evolution for time $nT$ and results in a power-$n$ decay of 
$\langle e_1| \psi \rangle/\langle e_0 | \psi\rangle.$ 
This observation remains true in the full Hilbert space with almost no modification.

Another way to obtain the power-$n$ decay of other eigenstates is to use a proper 
{\it window function} \cite{harris}.  Instead of performing FT as in (\ref{eq:phie}), 
one calculates
\begin{equation}
 |\phi_T^w(E)\rangle = \frac{1}{T}\! \int_{-T/2}^{T/2}{\!\!\!\!{\mathrm e}^{\frac{i E t}{\hbar}}|\psi(t)\rangle w_T(t) dt} 
\label{eq:phiw}
\end{equation}
with $w_T(t)=0.5\left(1-\cos\left(\frac{2 \pi t}{T}\right)\right), $ called a Hahn window, 
being one possibility.  This window gives a $1/T^2$ convergence (instead of $1/T$ without
a window) of $\beta(T).$ 

Further possible improvements will be apparent when discussing specific examples below.

\section{Bose-Hubbard model}

We now illustrate the method using the 
Bose-Hubbard (BH) model \cite{fisher89,jaksch98} as 
realized in a~gas of ultracold atoms in an optical lattice.
We are aware of limitations of the tight-binding Bose Hubbard description of the system for deep
optical lattices \cite{dutta11,sengst12}. The BH model provides, however, 
a realistic  and commonly used simplification of the problem.
A pioneer experiment~\cite{greiner02}
 observed a~quantum phase transition (QPT) from a~superfluid (SF)
phase to a~Mott insulator (MI), and stimulated research on
interacting many-body systems with external perturbations, for a~review see \cite{lewen2007}. 
A similar QPT was observed in 
a 1D realization of the BH model \cite{stoeferle04}. 
A $^{87}$Rb condensate was loaded in a~deep two-dimensional  ''transverse'' optical lattice potential 
 realizing an array of one-dimensional atomic tubes. 
A shallower optical potential along the tubes coming from a~pair of counter-propagating beams 
 $V/E_R=s \sin^2(2\pi x/\lambda)$ (in units of the recoil energy $E_R=h^2/2m_{Rb}\lambda^2$) was superimposed
on the system.
After preparing the system at different potential depth, 
the lattice  was modulated periodically at different frequencies measuring the absorbed 
energy (for details see \cite{stoeferle04}). 
In another realization of this essentially 1D experiment, 
the effects created by disorder were analyzed \cite{fallani07}. We shall consider both these 
situations in the following.

\subsection{Periodic lattice oscillations}

Let us consider first the excitations created by a periodic modulation of the lattice. 
The problem has been addressed in \cite{kollath06} where the energy gain due to such a
modulation was studied using a 1D Bose-Hubbard model in a parabolic trap and TEBD for 
the time evolution, obtaining results which were in a qualitative agreement 
with the experimental spectra \cite{stoeferle04}. 
Recently, a novel extended study appeared \cite{huo11} which analyzes in detail 
different possible excitations in the same model using both t-DMRG and a perturbative approach. 
This later calculation considers smaller and simpler system so, for pedagogical reasons, 
we start our analysis for cases discussed in \cite{huo11}.

The 1D Bose-Hubbard Hamiltonian  reads:  
\begin{equation}
\!H\! =\! -J \sum_{j}^{M-1} b_j^\dagger b_{j+1} + h.c. +\! \frac{U}{2} \sum_{j}^M n_j
\left(n_j - 1 \right) +\! \sum_{j}^M \epsilon_j n_j, 
\end{equation}
where $M$ is the number of sites, $b_j$ ($b_j^\dagger$) is the destruction (creation) operator of an 
atom at site $j,$
$n_j=b_j^\dagger b_j$ the number operator. The tunneling amplitude $J$ and the 
interaction energy $U$ depend on the lattice depth $s$ (see below),
$\epsilon_j=c(j-r_0)^2$ is the energy offset of a~given site, due to the harmonic trap
with $r_0$  the position of its minimum. When the lattice depth is modulated, 
$s(t)=s_0[1+\delta \cos(\omega t)],$ the energy absorption is enhanced at frequencies leading
to resonant excitation of some excited states.

The time-dependence of  $s(t)$ maps onto the corresponding time dependence of the tunneling amplitude
$J(t)$ and interaction strength $U(t)$ using, as in \cite{kollath06,huo11}, 
the approximate formulae due to Zwerger \cite{zwerger03}:
\be 
J(t)=\frac{4}{\sqrt{\pi}}s(t)^{\frac{3}{4}}\exp(-2\sqrt{s(t)})
 \ee
and
\be U(t) = 4\sqrt{2\pi} \frac{a_s}{\lambda}s^{1/4}(t)s^{1/2}_{\perp} \ee
where recoil energy units are used and $s_{\perp}$ is the depth of ``transverse'' potential 
creating the 1D tubes. 
Following \cite{huo11}, we denote by $U_0$ the value of $U$ in the absence of modulation ($\delta=0$) and by $J_0$ 
the corresponding tunneling amplitude. 
Also, we take $s_{\perp}=30$, $a_s=5.45$nm, $\lambda=825$nm and the curvature of the harmonic trap  
$c=0.0123$ as in \cite{huo11}. 

Consider first a system of  $N=36$ particles (corresponding to Fig.~2 of \cite{huo11}) 
in a deep lattice with $s_0=15$. 
The corresponding characteristic interaction energy  
is $U_0\approx 0.714,$ while the tunneling amplitude is $J_0/U_0\approx 0.01$.

\begin{figure}
\begin{center}
\includegraphics[width=0.9\columnwidth]{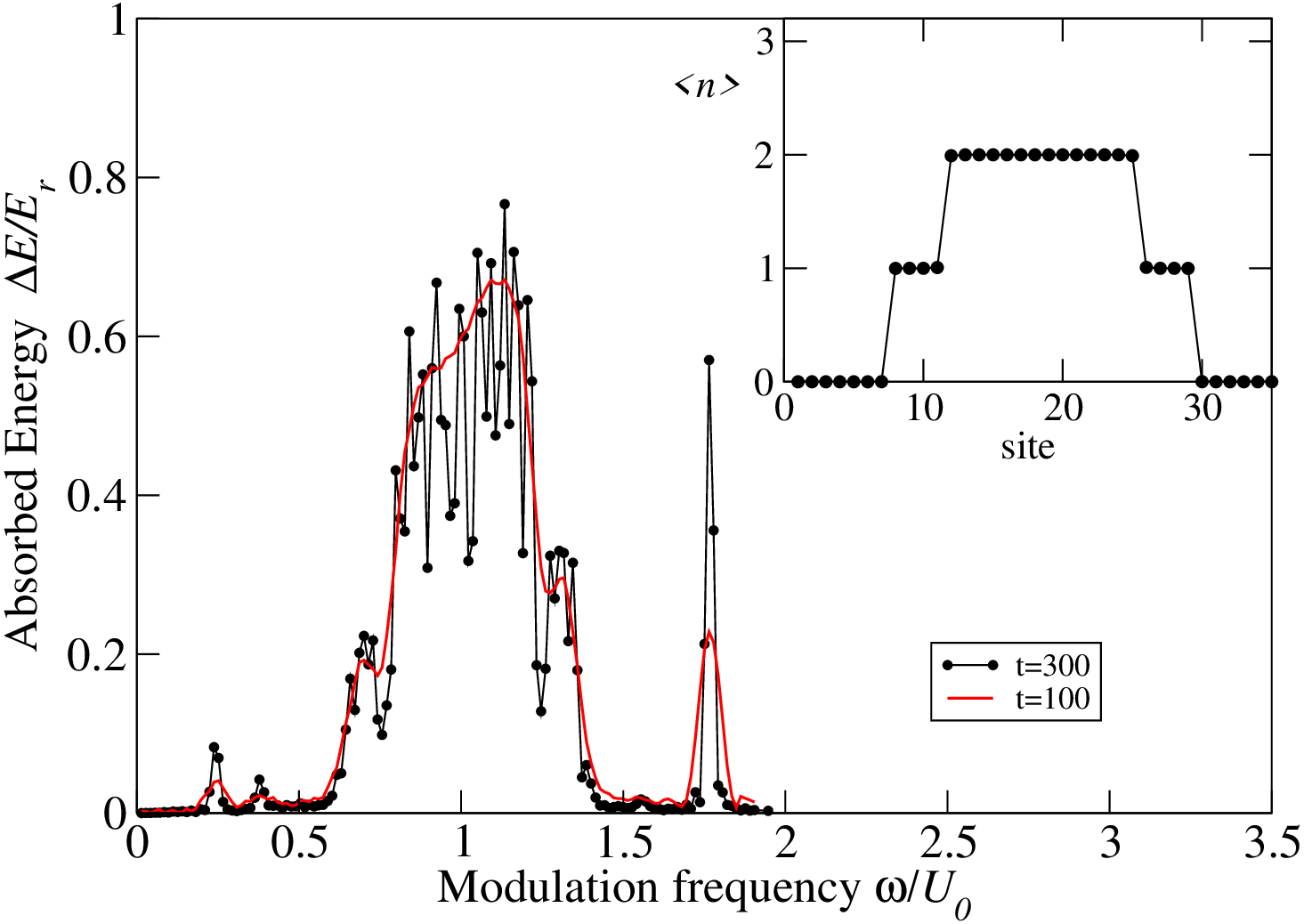}
\end{center}
\caption{(color online) Absorption spectrum  obtained for a 8\% modulation (i.e. $\delta=0.08$), 
after integration time $t=100$ (smooth red line) and $t=300$
(filled circles connected by the black line). The data for $t=100$ are multiplied by three.
The higher resolution obtained for longer times allows partially to separate various components 
due to individual excited states. Circles correspond to numerical data, black line 
connects them to guide the eye. The inset shows the average occupation numbers of the 
initial ground state of the system. 
}
\label{fig:ts3c}
\end{figure}

For any state in MPS representation - be it the ground state, an excited state
or a wavepacket - is it easy to compute expectation values of simple operators,
such as the average occupation number $\langle n_l \rangle$ at site $l,$ as well as its 
standard deviation $\Delta n_l=\sqrt{\langle n_l^2\rangle-\langle
n_l\rangle^2}.$
The ground state of the system is obtained by the TEBD algorithm with imaginary time evolution.
The average occupation number of sites, displayed in the inset of Fig.~\ref{fig:ts3c},
shows two Mott plateaus
with single and double occupancy. By simulating a periodic driving of the lattice using a real time TEBD 
evolution, we find the absorption spectrum. We take the modulation to be 8\%, that is
8 times bigger than in \cite{huo11};  this creates bigger excitation in the system, making 
extraction of the excited states easier. 
For the same purpose, we shift slightly the lattice and the bottom of the harmonic trap 
taking $r_0=18.62345$. This breaks the reflection symmetry of the system with respect to 
the center of the trap, making the analysis and  visualization of the excitations easier.  
Still, the spectrum for shorter evolution time, $t=100$ (for $\hbar=1,$ 
the unit of time is the inverse of the recoil energy) 
resembles the perturbative theory curve in Fig.~2 of \cite{huo11}. 
Longer integration time $t=300$ gives a better frequency resolution and several sharp peaks emerge.
The absorption for time $t=100$ multiplied by a factor of 3 seems to reproduce quite nicely
the global amplitude of the broad structure around $\omega=U_0,$ showing that, for such times,
absorption in this frequency region is linear. This is not true for isolated peaks at low 
frequency as well as for the peak around $\omega=1.75U_0,$ 
suggesting that these peaks correspond to isolated single levels 
(with the corresponding Rabi-type evolution). 
The integration is performed with time step $\Delta t=0.02$ using a standard second order 
Trotter approach assuming maximal occupation at the sites $i_{max}=6$ and $\chi=50$ 
(those values are highly sufficient for such a deep lattice and mean occupation of sites less than four).

The absorption spectrum now serves as a guide for extracting the excited states. Suppose we are interested in excited states contributing to the first small peak around $\omega/U_0=0.24$.
We take the wavepacket obtained at $t=300$ with the modulation at frequency $\omega/U_0=0.24$  as a starting wavepacket for the analysis. 
We first evolve this wavepacket for $T=10000$ with  time step $\Delta t=0.05$
using a {\it time independent} Hamiltonian (i.e. $\delta=0$) and find the autocorrelation
function $\tilde C_T(E)$.  
The latter shows a doublet 
around energy $E/U_0\approx 0.24,$ corresponding to two local excitations, one on the left side and one 
on the right side of the trap center.  
From this autocorrelation function, we obtain accurate eigenenergies.
In order to extract the eigenstate with the chosen eigenenergy $E,$  
we perform a Fourier transform on MPS states during a time sufficiently long
for all states outside the peak at $E/U_0\approx 0.24$ to be significantly damped (in this specific
case, $T_1=3000$ is convenient). 
This provides us with a  wavepacket $|\psi_1\rangle$ 
with a large overlap with the desired eigenstate (typically 0.7-0.9,
while the overlap of the eigenstate with the initial wavepacket was in the range 0.0001-0.01). 
The last step is to ``purify'' the wavepacket  $|\psi_1\rangle$ recursively:
the wavepacket $|\psi_i\rangle$ is propagated for a relatively long time ($T_i=6000$ in our specific case).
The Fourier transform of the time series at energy $E$ again which yields a new
wavepacket $|\psi_{i+1}\rangle,$ where the weight of all other
eigenstates has been decreased (a Hahn window is used to speed up this decrease).
This leads to an exponentially fast convergence of the wavepacket $|\psi_i\rangle$
towards the desired eigenstate.
We observed that numerical inaccuracies may cap the exponential convergence of this procedure after 
15-30 steps. In all propagating steps, we have again used time step $\Delta t=0.05$. 
In general, exact computational setup is adjusted taking into account the initial overlap, details of the spectrum and numerical stability.

Altogether, one iteration takes about 4 hours, using  8 threads on 4 core Intel i7 processors  
(the program is multi-threaded). We have performed 6 iterations to check the accuracy; a single additional iteration suffices for a 1\% accuracy on the occupation numbers.

\begin{figure}
\begin{center}
\includegraphics[width=0.9\columnwidth]{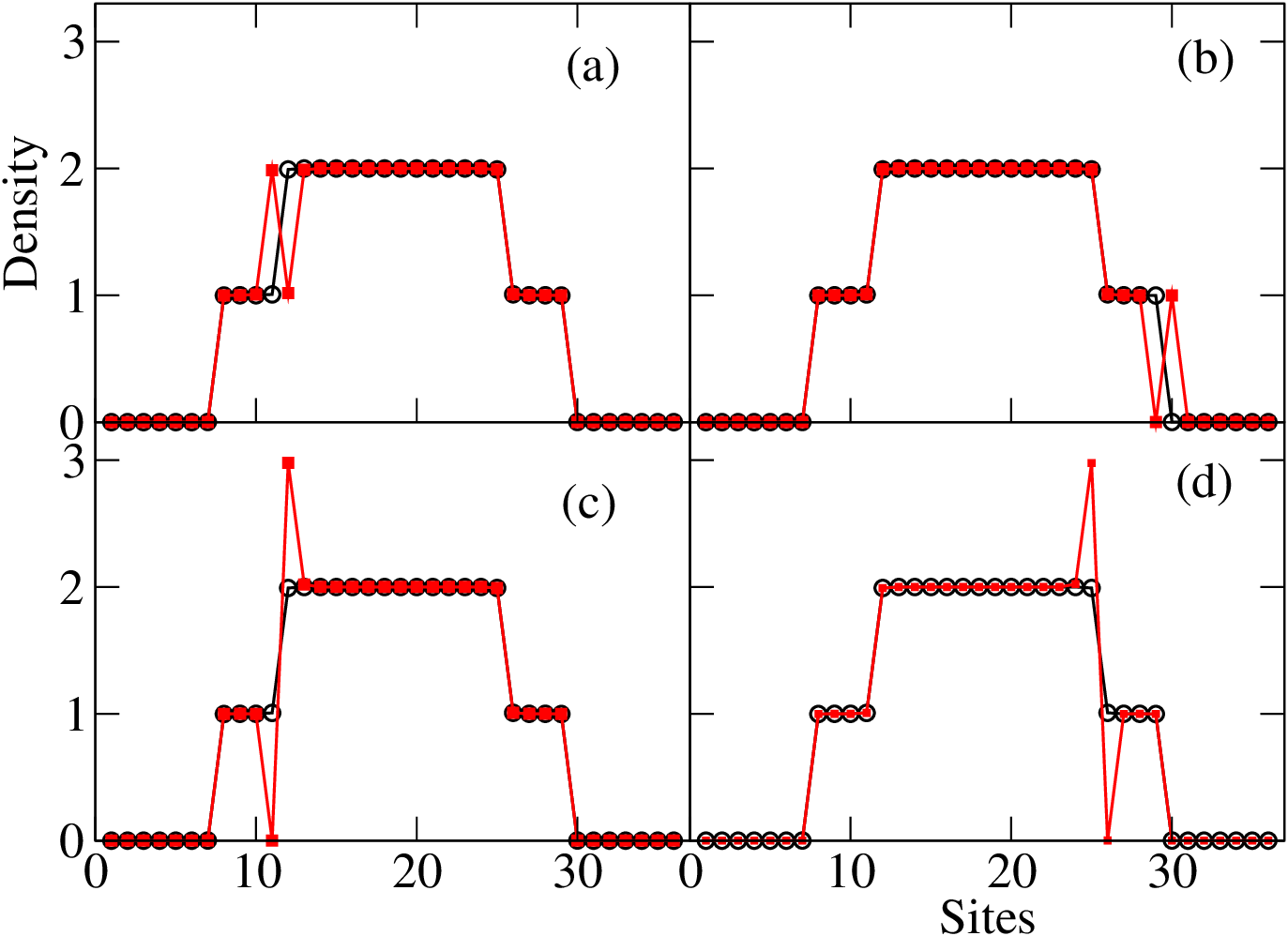}
\end{center}
\caption{(color online) 
Average occupation numbers per site for some selected excited states of the system 
analyzed in figure~\ref{fig:ts3c}. 
The black line with circles is the ground state, while red lines with squares show selected excited states. 
One of the two states corresponding to the absorption peak at $\omega/U_0=0.24$ is shown in panel (a)
(the other state is symmetric w.r.t. the trap center). 
Panel (b) gives the density for one of the states associated to the very small peak around 
$\omega/U_0=0.37$. Panels (c) and (d) present two states associated to the ``high'' frequency peak 
at $\omega/U_0=1.75$.
}
\label{fig:exc1}
\end{figure}

As mentioned above, two closely spaced in energy states correspond to the absorption peak
at $\omega/U_0=0.24$, they differ from the ground state by the hopping of one particle 
between two Mott domains with $n=1$ and $n=2$ occupations 
occurring either on the left or on the right side of the trap center. 
Had the trap be placed symmetrically with respect to the optical lattice,
the two elementary excitations would have been degenerate and the evolution from a symmetric 
ground state under a symmetric Hamiltonian would yield the symmetric combination of two excitations. 
A system with broken mirror symmetry allows for separation of these excitations,
the left one is shown in Fig.~\ref{fig:exc1}a. Fig.~\ref{fig:exc1}b shows a similar excitation 
where a particle hops from an occupied to an empty side (this time shown for a ``right'' excitation). 
The corresponding energy is larger since such a situation occurs further from the center, where the harmonic 
trap is steeper. Two other excitations shown in 
Fig.~\ref{fig:exc1}c and Fig.~\ref{fig:exc1}d are the ``left'' and ``right'' excitations associated 
to the high frequency peak at $\omega/U_0=1.75$ and correspond to a particle hopping from 
single to doubly occupied Mott zone creating a site with triple occupancy.

\begin{figure}
\begin{center}
\includegraphics[width=0.9\columnwidth]{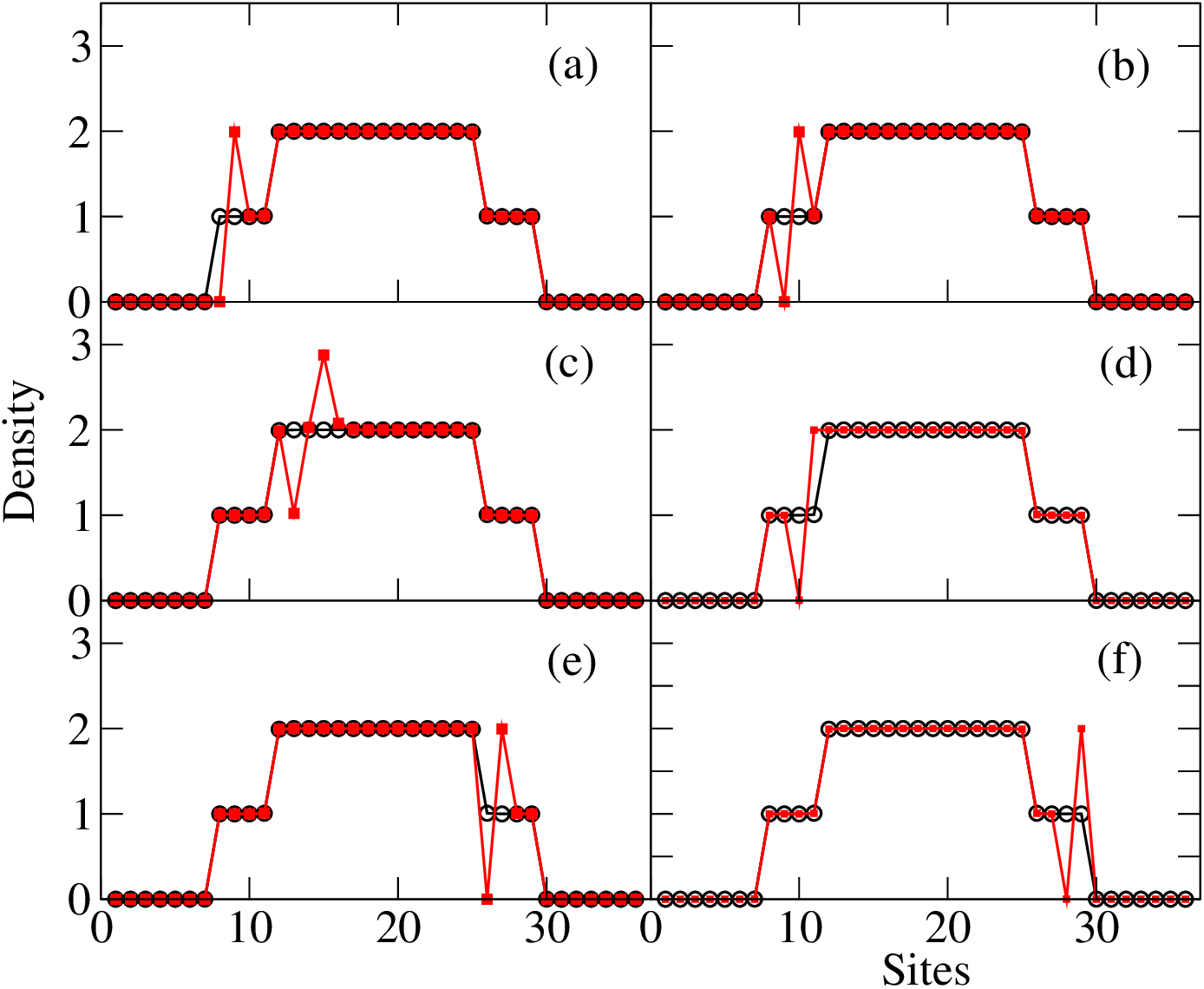}
\end{center}
\caption{(color online) 
Average occupation numbers for another set of excited states of the system  analyzed in Fig.~\ref{fig:ts3c}. 
The black line with circles represent the ground state while red lines with squares show 
selected excited states. 
}
\label{fig:exc2}
\end{figure}

All these excitations were correctly identified in \cite{huo11}, based on energy considerations 
(possible for simple situations occurring in the deep insulating regime for
large lattice depths). Our excited states provide a direct verification.
Similarly, we have checked that the broad structure around $\omega/U_0=1$ 
(in fact consisting of several distinct peaks as visualized by $t=300$ absorption spectra in 
Fig.~\ref{fig:ts3c}) corresponds to single particle-hole excitations in the central Mott domain 
as identified by \cite{huo11}.

Fig.~\ref{fig:exc2} presents another set of excited states. Fig.~\ref{fig:exc2}c resembles a typical
particle hole excitation in the central Mott domain, but not between neighboring sites. 
Thus, the corresponding excitation energy lies not in the broad structure 
but on its wings. All other plots in this figure present excited states 
not analyzed in \cite{huo11}. They correspond to absorption peaks occurring on the left 
shoulder (a-c) and on the right (d-f) shoulder of the central structure. 
They show either  particle-hole excitations
occurring in the lower $n=1$ Mott domain or more complicated situations.

It is apparent that our method provides reliably excited states in the localized regime. 
All these excited states are well represented by an MPS with relatively small $\chi$.
To discuss a more challenging situation, we move to another example, 
the absorption spectra studied in \cite{kollath06} where, due to a shallower optical lattice,
some excitations may occur in the superfluid regions. 
In particular, two structures in the absorption spectra  
around $\omega/U_0\approx 1$ as well as $\omega/U_0\approx 2-2.3$ were observed, 
in agreement with the experiment \cite{stoeferle04}.
The former resonance corresponds to the gap in the MI phase 
(i.e. to the particle-hole excitation in the Mott phase). Its appearance was interpreted as an 
evidence for the presence of the Mott phase. The latter has no such easy explanation; 
it was argued heuristically \cite{kollath06} that 
it corresponds to a particle-hole excitation in the superfluid component. 

\begin{figure}
\begin{center}
\includegraphics[width=0.9\columnwidth]{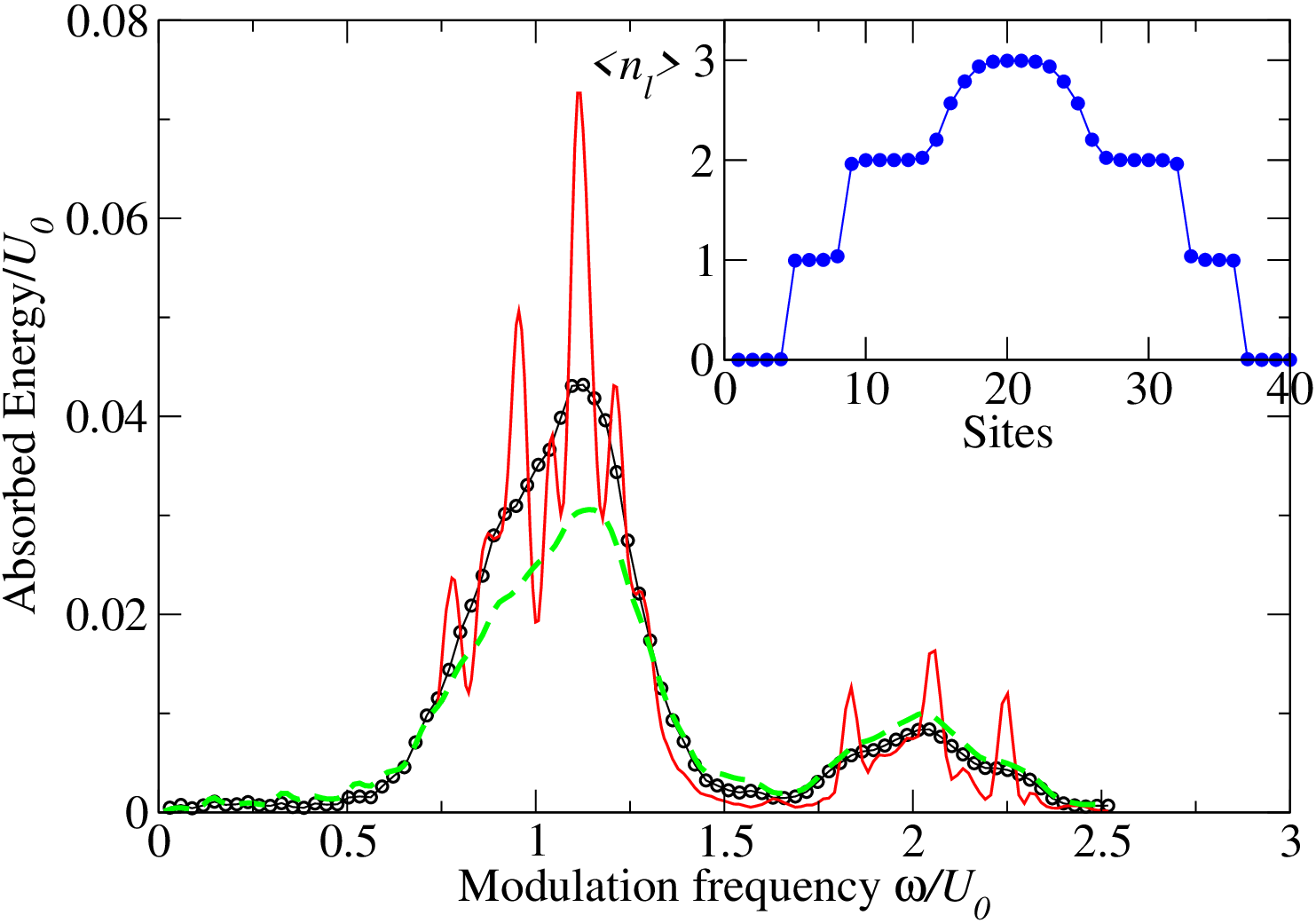}
\end{center}
\caption{(color online) 
Absorption spectrum for a system of 65 particles on 40 sites, in an 
harmonic trap with curvature $c=0.006,$ centered around $r_0=20.62345$. 
The mean lattice depth is $s_0=12$ (corresponding to $U_0=0.6752$ recoil energy). 
It is modulated sinusoidally with 
1.67\% relative amplitude. The smooth black line with open circles correspond 
to a modulation duration $t=50/E_R$; for longer modulation time $t=200/E_R,$ sharper peaks 
become resolved (red thicker line). 
The absorbed energy is divided by 4 in the later case, 
to compensate for a longer perturbation. 
The thick (green online) dashed line shows the absorption curve for 
a much stronger 20\% modulation for $t=50/E_R$, rescaled by the intensity factor $12^2$.}
\label{fig:akol}
\end{figure}

Our method allows to test the correctness of this claim. We take parameters
of Ref.~\cite{kollath06} with a slight modification. 
As in the previous example, we break the symmetry of the problem by shifting the lattice 
minima with respect to the harmonic trap. 
Such a situation seems generic from experimental point of view and 
allows to separate individual excitations. 
The absorption spectrum, similar to that of Fig.~4 of Ref.~\cite{kollath06},
is shown in Fig.~\ref{fig:akol}.
The initial state - ground state at $s_0=12$ - is excited 
by a $\delta=0.2$ (20\%) modulation   for a duration $t=50/E_R$ - the result is shown as a thick dashed line. 
We observe a broad peak around modulation frequency $\omega/U_0=1$ as well as a much smaller broad peak 
around $\omega/U_0=2$ 
\footnote{In \cite{kollath06}, the relative amplitude of these two peaks differs from our result. 
We have tried to reproduce their curve also for precisely the same situation, i.e. a with $r_0=20.5,$
the symmetric case, but still the discrepancy persisted.}. 
Fig.~\ref{fig:akol} shows the absorption spectra for a more gentle
driving, i.e. $1.67\%$ modulation. For an easier comparison, the strong modulation curve 
is rescaled by $12^2$ factor (square of the relative amplitude of modulation). 
The comparison of the two curves show that the structure around $\omega/U_0=2$ behaves perturbatively, 
while the broad resonance around $\omega/U_0=1$ shows signature of saturation. 

The wavepacket obtained for a given frequency, at the end of the modulation, 
is a starting wavepacket for our procedure. 
The strong 20\% modulation, as used in~ \cite{kollath06}, 
leads to a ''too excited'' wavepacket for the TEBD algorithm to work effectively over the long time 
required for a high-resolution FT (see, however, a discussion below). 
Of course, the excited states do not depend on the strength of the modulation 
(while the shape of the absorption curve in the nonperturbative regime does). 
For that reason, we use the wavepacket obtained after a ``gentle'' modulation 
with a smaller (1.67\%) amplitude.    

Fig.~\ref{fig:akol} shows also the absorption curve for a longer modulation interval $t=200/E_R$ 
which allows for a partial resolution of broad peaks. In particular, the structure around  
$\omega/U_0=2$ shows narrower partially resolved peaks 
around $\omega/U_0=1.84,\ 2.06,\ 2.25$.

\begin{figure}
\begin{center}
\includegraphics[width=0.9\columnwidth]{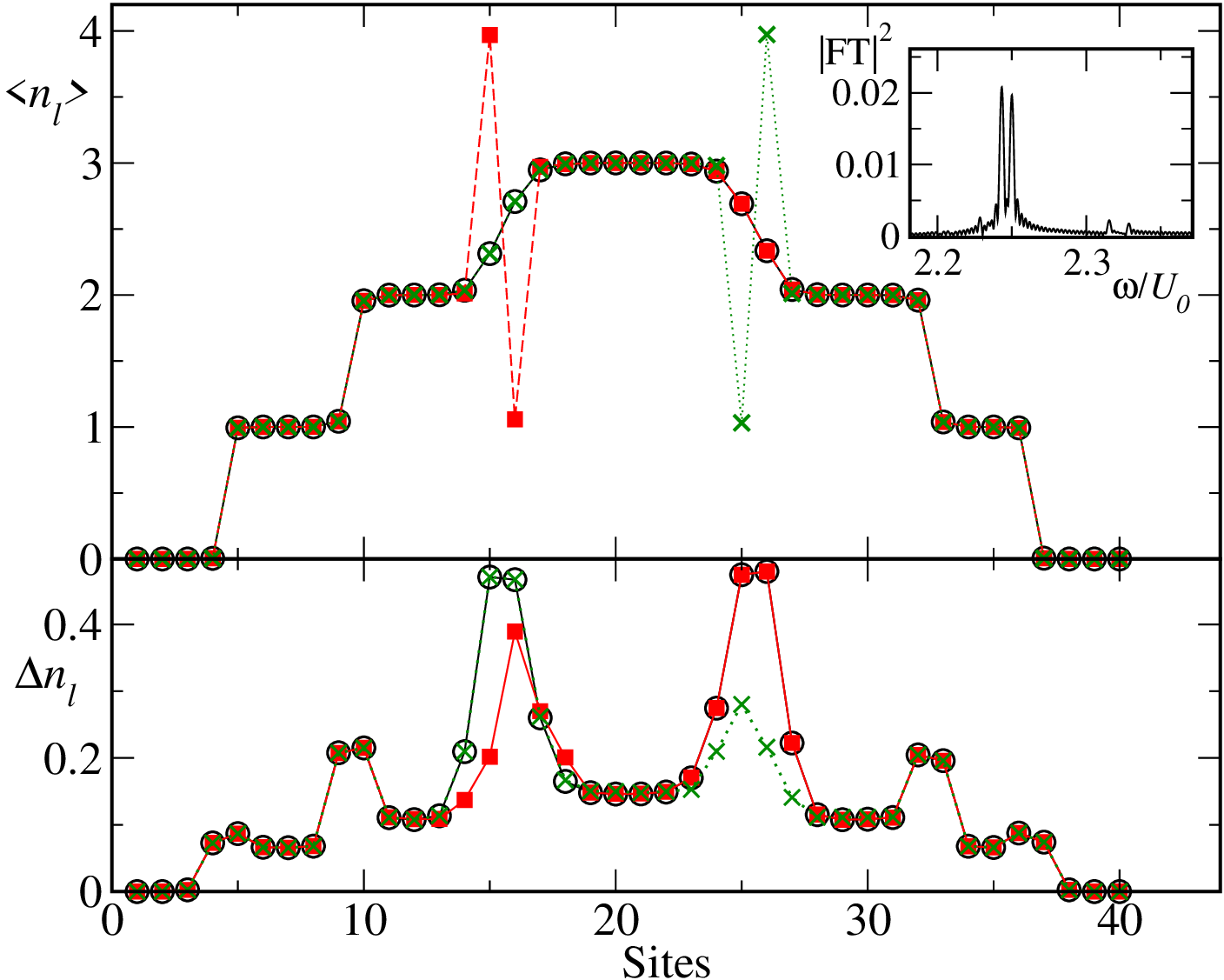}
\end{center}
\caption{(color online) 
Black circles: average occupation number for the ground state of 65 particles in a lattice with depth 
$s=12$ (40 sites) and an harmonic trap with curvature $c=0.006,$ centered around $r_0=20.62345$  
(to  separate left and right excited states w.r.t. to the trap center). 
The filled squares (red online) connected by a dashed line and the crosses (green online) 
connected by a dotted line show occupation numbers for two extracted excited states  
showing a particle-hole excitation in the SF region. 
The excess energies (energies measured with respect to the ground state energy) of the states are 
 $\Delta E/U_0=2.2502$ 
(resp. $\Delta E/U_0=2.2434$) for the ``red squares'' (resp. ``green crosses) in excellent agreement with
the  positions of two closely spaced peaks in the FT of the autocorrelation function
shown in the inset. 
The lower panel shows the standard deviation of the occupation number on each site, for the ground state 
(open circles) and the excited states, the latter shows reduced fluctuations. }
\label{fig:koll}
\end{figure}

Let us concentrate first on the wavepacket modulated with frequency $\omega/U_0=2.3$. 
We perform a FT of the autocorrelation function over a long time and observe several partially 
overlapping peaks. As shown in the inset of fig.~\ref{fig:koll}, 
we identify a doublet of states, relatively well separated from the others, significantly
excited from the ground state by the modulation. Fig.~\ref{fig:koll} shows the average occupation numbers
$\langle n_l\rangle$ at
site $l$ for the two excited states, extracted by our method, together with the ground state.
Observe that, indeed, these states correspond to a particle-hole excitation in the SF regime 
(with average occupation numbers between 2 and 3) confirming the  suggestion of~\cite{kollath06}. 
The two states are quasi-symmetric images; they have slightly different energies, 
because of the choice $r_0=20.62345$ which breaks parity w.r.t. the trap center. 

We have found that, in order to obtain a fully converged excited state 
(e.g. the excitation 
appearing on the right hand side of the center of the trap represented as crosses in Fig.~\ref{fig:koll}), 
another technique is useful. Knowing from the preliminary runs where
the local (in space) excitation should be created, we prepare a new wavepacket by a local modulation, 
i.e. local chemical potential is modulated for a few nearby sites (around site 25 for this particular case). 
Explicitly $\mu_i(t)= \mu_i(1+\delta \exp[-(i-i_0)/w^2] \sin (\omega t))$
where $w$ is the spatial width of the modulation. 
Experimentally, such a modulation could be in principle created by an additional focused laser. 
Here it is used in the numerical simulation to speed up the convergence as well as to isolate excitations 
quasi-degenerate in energy but spatially well separated.

The data in Fig.~\ref{fig:koll} show some interesting differences when compared to the Mott
regime previously discussed.
While, in Figs.~\ref{fig:exc1}c and \ref{fig:exc1}d, the excited states
display a transfer of particles towards the trap center - explaining why they have
an excess energy (compared to the ground state) smaller than $2U_0$ -  it is the opposite
for states in  Fig.~\ref{fig:koll} where particles are transferred further from the trap center,
and the associated excess energy is larger than $2U_0.$
The MPS representation of the excited states makes it easy to calculate several observables, 
not only the average occupation number of sites. The lower plot in Fig.\ref{fig:koll} shows 
the standard deviation of the occupation of sites $\Delta n_l$, 
both for the ground state and the excited one. A small (resp. large) standard deviation is a signature
of a Mott insulator (resp. superfluid region). The excited state has lower  $\Delta n_l$ in the region of 
excitation, proving that it is more an ``excited Mott''. This is confirmed by the occupation numbers
themselves which are close to integer values for the excited state, with larger fractional part
for the ground state.

\begin{figure}
\begin{center}
\includegraphics[width=0.9\columnwidth]{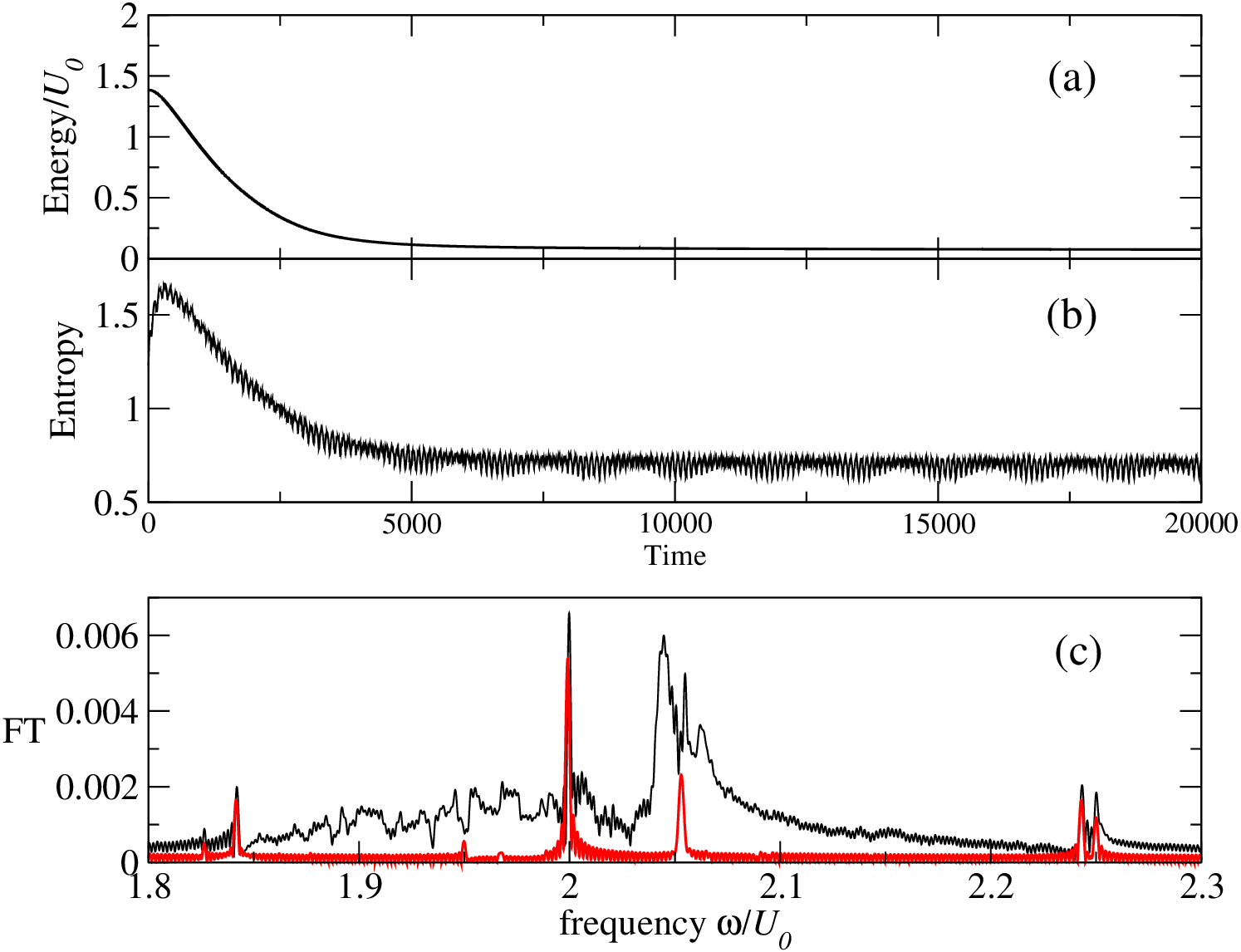}
\end{center}
\caption{(color online) 
Long time evolution of a wavepacket prepared by a modulation at frequency $\omega/U_0=2$ 
and then evolved with the TEBD algorithm. 
The excess energy over the ground state is plotted in top frame (a).  It is not conserved but,
after the initial drop, it stabilizes. Similar behavior occurs for the entanglement entropy (b), 
with additional
oscillations because the wavepacket is a sum of relatively few localized states.
The FT of the autocorrelation function (c) does not display any clear peaks if performed in the
[0,5000] time interval (black thin line), but definitely shows several
well defined excited states over the time interval [15000,20000] (thick line, red online).}
\label{fig:conver}
\end{figure}

To  explore other excited states contributing to the structure 
around $\omega/U_0=2,$ we prepare the wavepacket with modulation frequency $\omega/U_0=2$.  
The first attempt to calculate the autocorrelation function by propagating such an excited wavepacket 
with $\chi=70$ (and a local Hilbert space on each site containing occupation numbers
up to imax=7) failed, as e.g.
the energy in the system was not conserved.  
We chose, however, to simply run the algorithm for a much longer time and observed 
an unexpected behavior depicted in Fig.~\ref{fig:conver}. 

Fig.~\ref{fig:conver}a presents the energy of the wavepacket versus time 
(in recoil units, multiplication by $U_0=0.6752$ give time in units of $1/U_0$). 
After the initial decrease, we observe {\it stabilization} of the energy at some value 
above the ground state. Fig.~\ref{fig:conver}b presents the corresponding time dependence 
of the entropy of entanglement in the system defined as $S=\mathrm{max}_l\ S_l$ with $S_l$ given 
by Eq.~(\ref{eq:ent_ent}). It also decays and stabilizes.
This behavior can be understood by performing the FT of the wavepacket 
autocorrelation function (shown in Fig.~\ref{fig:conver}c). 
While the FT over the time interval [0,5000] displays  broad erratic structures,
the FT over the time interval [15000,20000]  gives sharp peaks at some frequencies. 
The heuristic understanding of this behavior is the following. 
The initial excited wavepacket is a linear combination of eigenstates of the system. 
Some of them are well represented as an MPS, some other ones
are represented poorly. During the time evolution the ``well represented'' part 
evolves accurately while the remaining part produces signatures of TEBD algorithm
breakdown. During the course of evolution, a {\it self-purification} of the wavepacket 
occurs with the non-converged fraction seemingly disappearing. 

The information on the non-converged fraction is actually lost in the TEBD evolution,
when some $\lambda_{\alpha}^{[l]}$ values are discarded (those with  $\alpha>\chi$.)
No, or little,  information is lost on the well represented excited states. 
Thus, unconverged TEBD time evolution filters out information 
which anyway cannot be obtained, purifying the wavepacket.

\begin{figure}
\begin{center}
\includegraphics[width=0.9\columnwidth]{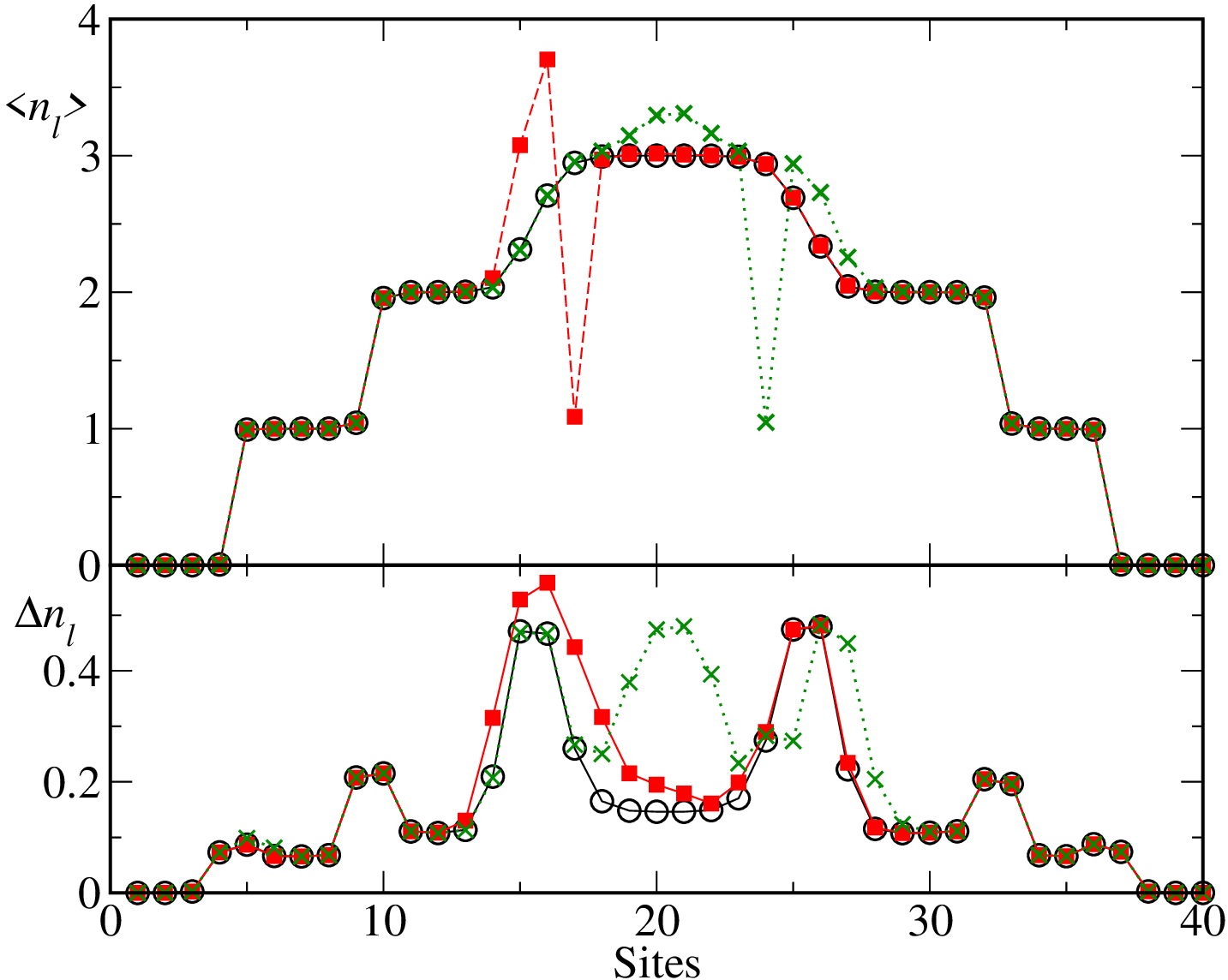}
\end{center}
\caption{(color online) 
Average occupation numbers of exemplary excited states showing 
a significant superfluid-type excitation, when compared to the ground state (black open circles). 
Red squares connected by the dashed line correspond to the excited state with 
excess energy $\Delta E/U_0=2.2418$  
(lower energy peak of the doublet in the inset of Fig.~\ref{fig:koll}). 
Green crosses connected by dotted lines correspond to a state 
(excess energy $\Delta E/U_0=1.9504$) with significant excitation around the trap center. 
The lower panel shows the standard deviation of the occupation of sites for 
the ground state (open circles) and two excited states. 
The latter show regions of increased superfluid-like fluctuations. 
}
\label{fig:kollsf}
\end{figure}

A further confirmation comes from the fact that  
the peaks of the FT  provide the energies of excited states which can now be extracted. 
Fig.\ref{fig:kollsf} show examples of excited states involving excitations in the superfluid regime. 
Contrary to the previous examples, these excited states 
show increased particle number fluctuations, see lower panel of Fig.\ref{fig:kollsf}.
 
 \begin{figure}
\begin{center}
\includegraphics[width=0.9\columnwidth]{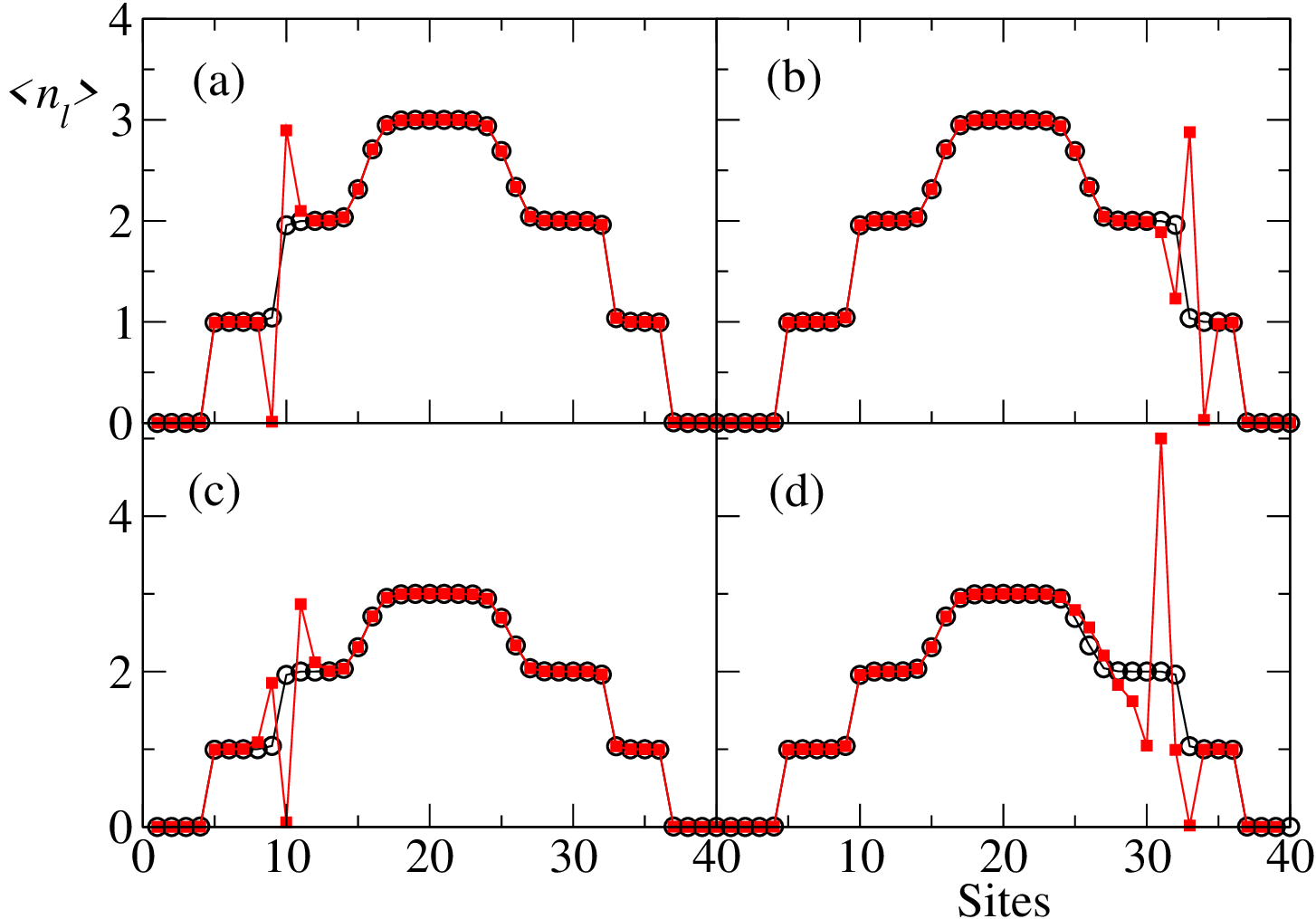}
\end{center}
\caption{(color online) 
Average occupation numbers of exemplary excited states showing 
localized particle-hole excitations, compared to the ground state (black open circles). 
Red squares connected by dashed lines correspond to the excited states. 
(a) Excitation similar to those discussed for the smaller system, i.e. particle-hole excitation 
on the transition region between two Mott plateaus (excess energy $\Delta E/U_0=1.8417$). (b) and (c) 
show two particles being transfered between sites, the excess energies are $\Delta E/U_0=1.9988$ and
$\Delta E/U_0=2.0539$. (d) gives an example of higher lying excited state $\Delta E/U_0=7.5280$ with 
a partially melted, right $\langle n_l \rangle=2$ Mott plateau. 
}
\label{fig:kollmo}
\end{figure}
\begin{figure}
\begin{center}
\includegraphics[width=0.9\columnwidth]{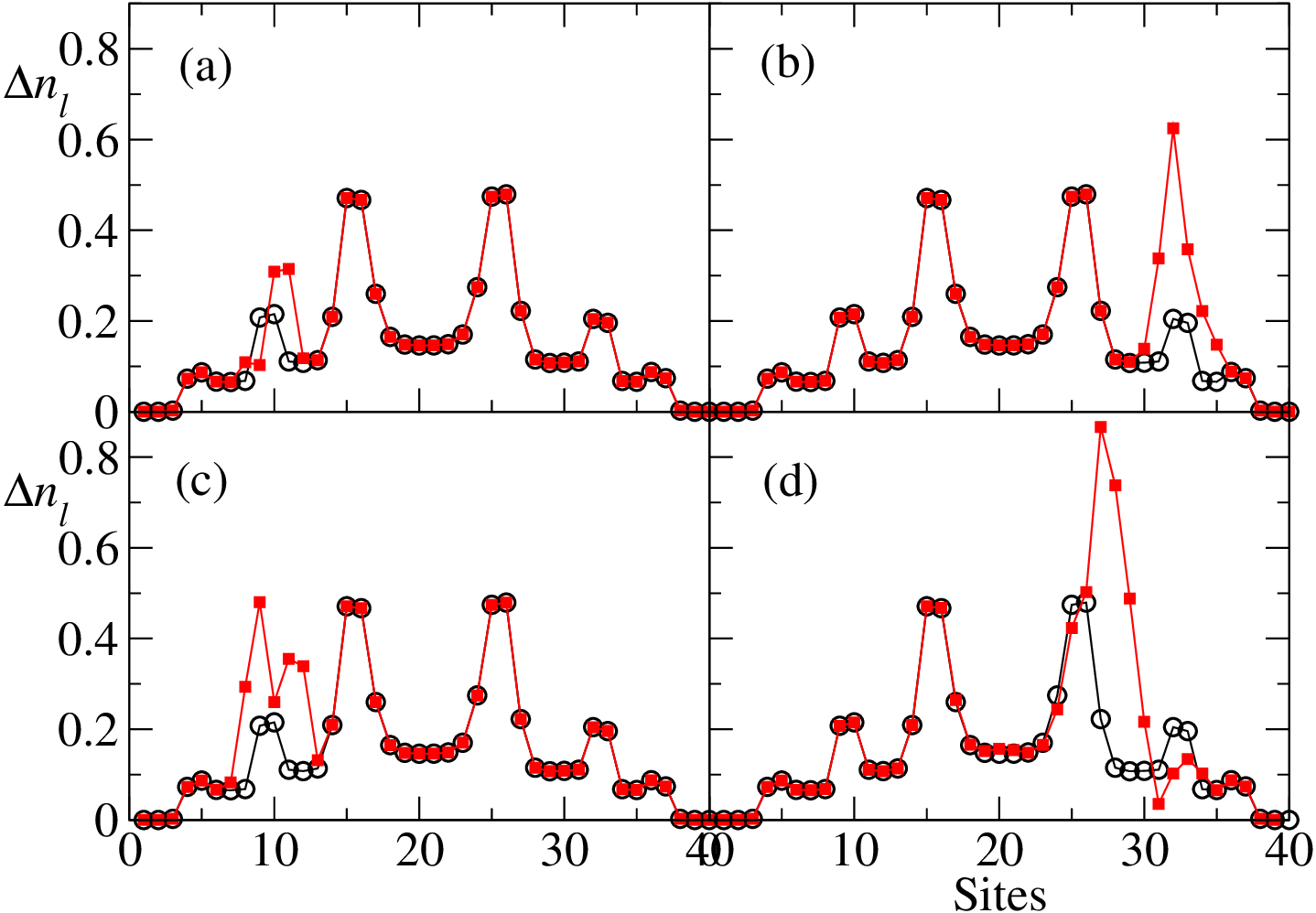}
\end{center}
\caption{(color online) 
Standard deviations of the occupations number, for the states presented in Fig.~\ref{fig:kollmo}.
Local excitations lead to an increase of particle number fluctuations.
}
\label{fig:kollmosd}
\end{figure}

States presented in Fig.~\ref{fig:kollmo}a--Fig.~\ref{fig:kollmo}c are other examples, 
well localized on the edge between the $\langle n_l \rangle=1$ and $\langle n_l \rangle=2$ zones. 
The corresponding standard deviations shown in Fig.~\ref{fig:kollmosd} show that these excitations, 
nevertheless, lead to an increase of local particle number fluctuations. 
Fig.~\ref{fig:kollmo}d shows a state with much larger excitation energy, 
almost $8U_0$. 
It shows that our method may produce states with quite high and complicated excitations. 
This state displays several particle transfers as well as almost a 
transfer of a part of the right hand side Mott  $\langle n_l \rangle=2$ plateau into a superfuid region.

In effect a vast majority of states identified by the FT of the autocorrelation function
can be extracted. Naturally, 
most of these states correspond to local excitations and not too high entanglement entropy.

\subsection{Florence experiment}

A second important example is the 1D experiment of the Florence group \cite{fallani07} where a second,
weaker lattice potential (incommensurate with the primary lattice) is added to simulate a disorder. 
Its strength   is denoted by $s_2$. Theoretically,
the addition of a~disordered potential reveals a~new insulating, compressible and gapless phase, called
Bose glass (BG) \cite{fisher89}. Experimental results confirm its existence~\cite{fallani07,white09}, 
but are not fully conclusive: the system is 
prepared from a~low temperature BEC by ramping up
an optical lattice, starting from a~SF initial state. Preparing a~MI or a~BG requires, 
generally non-adiabatic, 
passage through a~quantum phase transition~\cite{dziarmaga02,polkovnikov05}, 
thereby exciting the system.

The Florence experiment is described in~\cite{fallani07}. 
We simulated the experiment as detailed in~\cite{zakrzewski09}: 
we integrate numerically the evolution of the system
when $s$~increases using 
the TEBD algorithm \cite{vidal03}, starting from the SF ground state at $s=4.$
At the same time, a secondary
shallow lattice creating the disorder is turned on. Its effect is to modify the energy offset of the harmonic wells
at different sites yielding now:
 \begin{equation}
\epsilon_j= c(j-r_0)^2 + s_2 E_{R2} \sin^2 \left(\frac{\pi j\lambda}{\lambda_2} + \phi
\right).
\label{trap}
\end{equation}
The curvature of the trap, $c$ depends on the harmonic trap and on the additional trapping 
potential created by the transverse profile of the laser beams \cite{fallani07}.
 
As $s$~increases, we found~\cite{zakrzewski09} that the overlap of the dynamically evolved 
wave packet on the ground state 
for that $s,$ rapidly decreases in the region of the SF-MI transition 
(around $s=8-9$ for the experimental parameters)
down to less than 10\% for the final $s=14,$ a signature of broken adiabaticity. 
The situation is even worse (with the final overlap of the order of $10^{-6}$) for the disordered system.
Disorder enriches the energy level structure around the ground
state, killing
adiabaticity  both for shallow  \cite{edwards08} and deep \cite{zakrzewski09} lattices.  

To identify the induced excitations, 
we compute the various excited states
building the dynamically created wavepacket, by evolving it further at the {\it constant} final $s$ value
and calculating the FT in Eqs.~(\ref{eq:ce},\ref{eq:phie}), as explained in the previous Section.  
In particular we characterize the states obtained using similar observables 
(easily obtained in the~MPS representation~\cite{mcculloch}):  the
average occupation number $\langle n_l\rangle$ at site $l$ and its standard deviation 
$\Delta n_l.$
As a~by-product, we can also read off the average number of particles in the 
left part of the system $N_l=\sum_{i=1\ldots l} n_i$
and its standard deviation $\Delta N_l.$ 
This quantity is relatively easy to measure in a~experiment. In our system
it qualitatively resembles the entanglement entropy, as shown below \cite{klich}.

\begin{figure}
\begin{center}
\includegraphics[width=0.9\columnwidth]{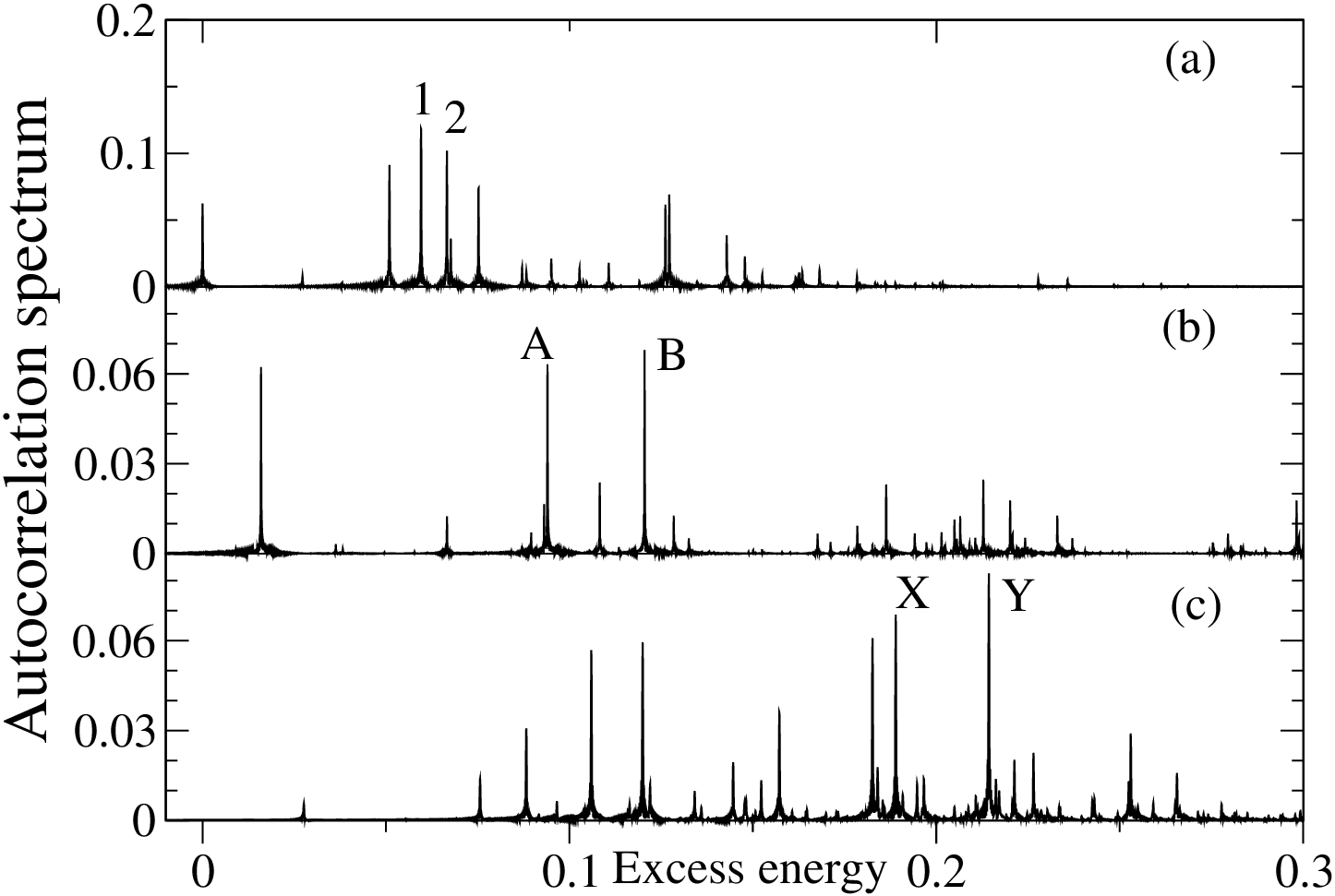}
\end{center}
\caption{Autocorrelation spectra, Eq.~(\ref{eq:ce}), obtained dynamically for $s$=14 
after switching on the optical lattice, for disorder strengths $s_2$= (a) 0, (b) 0.4375, (c) 2.1875. 
All parameters are taken to approximate the experimental situation \cite{fallani07}.
The energy levels of the system appear as peaks (origin at the zero-point energy), 
with height $|c_i|^2$  from eigenbasis expansion. Peak labels are for further reference.}
\label{fig:autocorr_spectra}
\end{figure}

\begin{figure}
\begin{center}
\includegraphics[width=1.0\columnwidth]{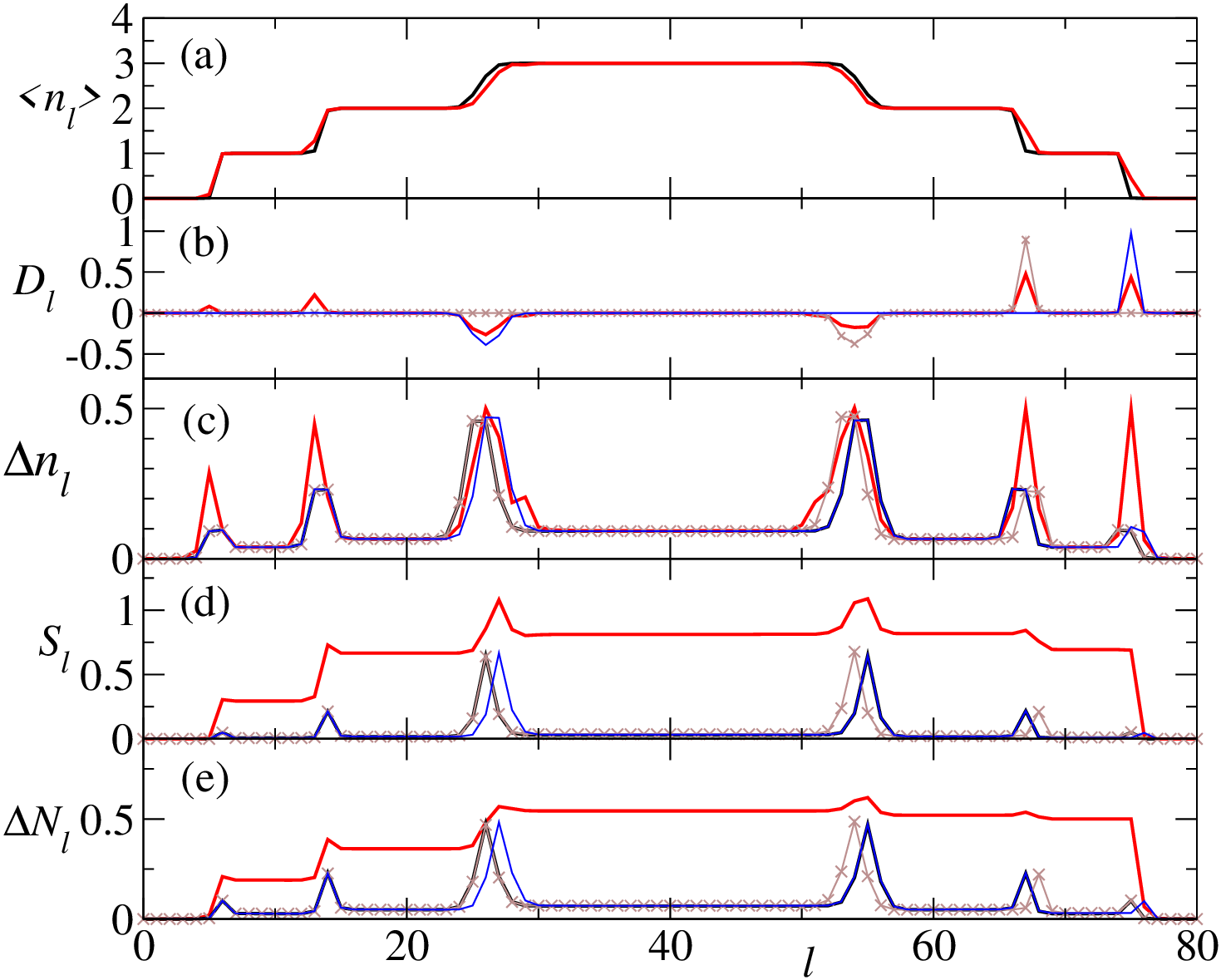}
\end{center}
\caption{(color online)
Properties of states of the trapped BEC in
a deep optical lattice ($s$=14),
without disorder ($s_2$=0), $r_0$=0.12345 in Eq.~(\ref{trap}) to break the parity symmetry w.r.t. 
the trap center.
The black line refers to the ground state, the red line to
the dynamically prepared wavepacket, the brown (with crosses) and blue lines
to the two excited states with the largest populations denoted as 1 and 2 in Fig.~\ref{fig:autocorr_spectra}a.
(a): Average occupation number $\langle n_l \rangle$ on each site.
(b): Difference $D_l=\langle n_l \rangle - \langle n_l \rangle_{\mathrm{Gnd\ state}}$ 
(c):
$\Delta n_l=\sqrt{\langle
n_l^2\rangle-\langle n_l\rangle^2}.$ (d): Entanglement entropy, Eq.~(\ref{eq:ent_ent}); almost zero in the Mott plateaus for eigenstates
- implying their approximate separability; large for the wavepacket.
(e): Variance of the number of atoms to the left of site $l$. It is akin to the entanglement
entropy, showing a~marked difference between the eigenstates and the wavepacket.
\label{fig:full_s0}}
\end{figure}

Fig.~\ref{fig:autocorr_spectra} shows the autocorrelation
spectrum, eq.~(\ref{eq:ce}), of the dynamically created wavepackets, at increasing disorder strengths.
In the absence of disorder ($s_2=0$), about ten states are significantly excited. 
In Fig.~\ref{fig:full_s0}, we show various
relevant quantities for the ground state, few excited states
and the wavepacket. The average occupation number $\langle n_l\rangle$ on each site $l$ has the
well known ``wedding cake'' structure, with large MI regions with integer  $\langle n_l\rangle$ separated
by narrow SF regions. Because the energy excess brought by non adiabatic preparation is small,
all significantly populated excited states have similar shapes. Clearly,
all excitations take place in or around the SF regions: these are transfers of one atom from the 
edge of a~Mott plateau to the edge of another Mott plateau or to the 
neighboring SF region (``melting'' of the Mott plateau).

The standard excitations in an homogeneous system (without trapping potential) 
such as a~particle-hole excitation
in a~Mott plateau are absent in the stationary states because they are dynamically unstable. 
They are also
absent in the wavepacket because they are energetically too costly: 
a~particle-hole excitation costs at least
one interaction energy $U\approx0.6E_R$, 
much larger than the excess energy of the wavepacket $0.1E_R$. 
Observe, therefore, that we are in a regime significantly different from the one
 discussed in the previous Section where excited states populated by 
modulation of the lattice depth have been considered. 
Here we are much ``closer'' in energy to the ground state.
  
The description used in~\cite{gerbier05} where the ground state is 
contaminated by local particle-hole excitations
is not compatible with our findings. 
A description in terms of melting of the MI ~\cite{gerbier07} seems more relevant.
The local occupation fluctuation $\Delta n_l$ confirms the existence of large
MI regions with low $\Delta n_l$ separated by SF peaks with larger $\Delta n_l.$ While the ground state
and excited states look very similar, the
wavepacket is different, with higher $\Delta n_l$ in the SF peaks. An even greater difference
is revealed by the entanglement entropy. For the eigenstates, it is essentially zero
in the MI regions and displays sharp peaks in the 
SF regions. In stark contrast, the $S_l$ of the wavepacket is 
also non-zero in the Mott plateaus. Similarly, $\Delta N_l$ is much larger in the Mott plateaus for the
wavepacket than for the eigenstates. This suggests the existence of long-range entanglement 
between SF interfaces linking different
MI regions caused by long range particle dislocations, absent in the eigenstates.
 
\begin{figure}
\begin{center}
\includegraphics[width=1.0\columnwidth]{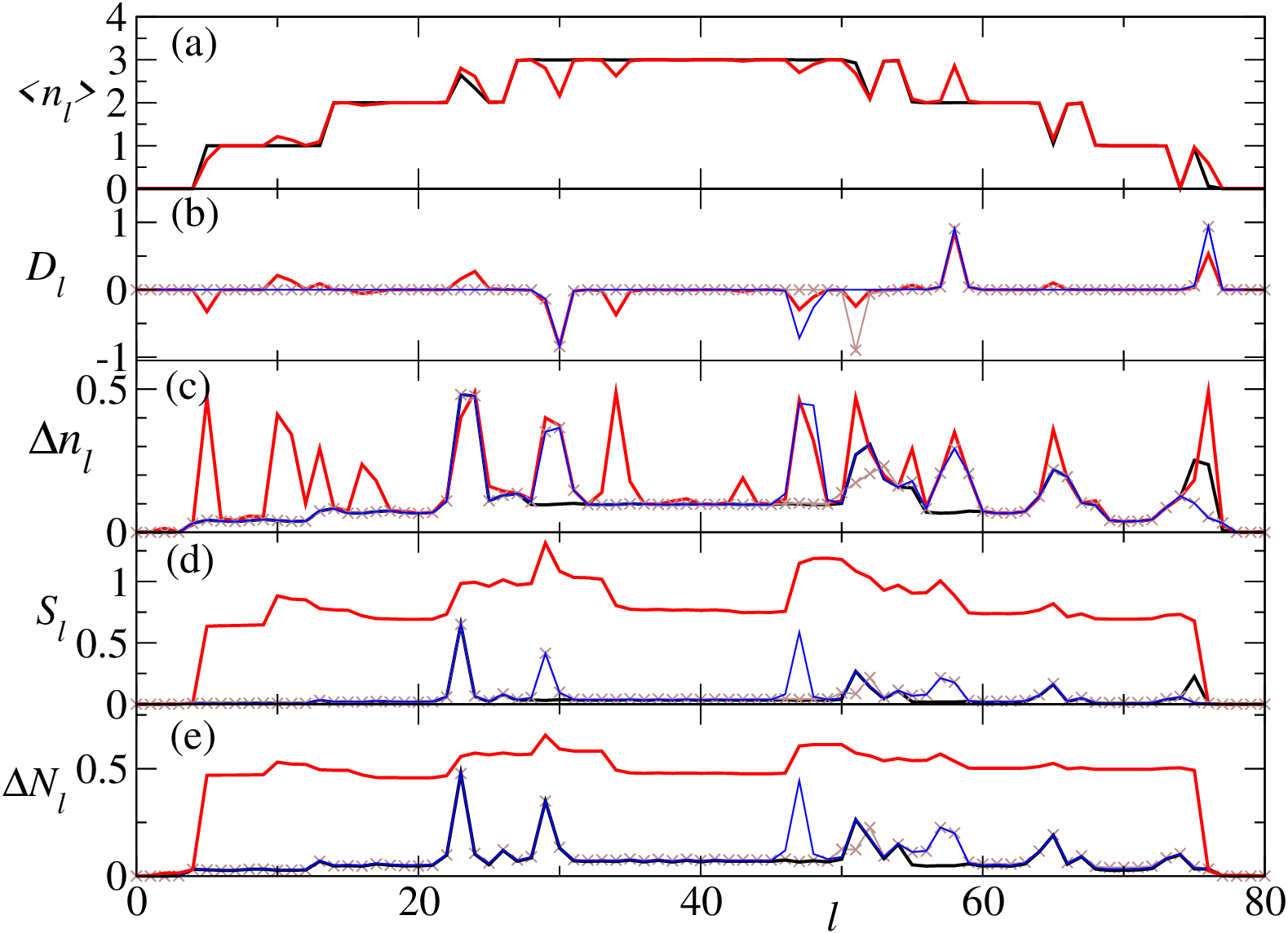}
\end{center}
\caption{(color online) Same as Fig.~\ref{fig:full_s0} but for
 small disorder ($s_2$=0.4375, $r_0$=0 and $\phi$=0.5432 in Eq.(\ref{trap})).
Brown (blue) lines correspond to excitations A and B indicated in,
Fig.~\ref{fig:autocorr_spectra}(b).
Mott insulating regions are clearly visible for all eigenstates, more vague for the wavepacket. 
\label{fig:full_s05}}
\end{figure}

Disorder strongly modifies the properties of the system. 
Possible phases are analyzed in details in \cite{Roscilde08}.
We consider first a~small disorder $s_2$=0.4375.
The breakdown of adiabaticity
is stronger in presence of disorder, 
with twice larger excess energy - still below the particle-hole excitation energy  -
and more states significantly excited.
The properties of various states are shown for small $s_2$ in Fig.~\ref{fig:full_s05}.
Several MI regions - identified by plateaus in $\langle n_l\rangle$ - still exist, separated by intermediate
parts which are either SF
or a~BG. 
Remarkably, the ground state and the excited states do have very similar structures, 
with several visible Mott plateaus.
In contrast, these plateaus are less visible (only the central one seems to survive, 
with a~reduced size) for the wavepacket.
This can be interpreted as a~partial melting of the BG and MI phases, 
producing a~thermal insulator \cite{gerbier07}.

\begin{figure}[ht]
\begin{center}
\vskip 1truecm
\includegraphics[width=1.0\columnwidth]{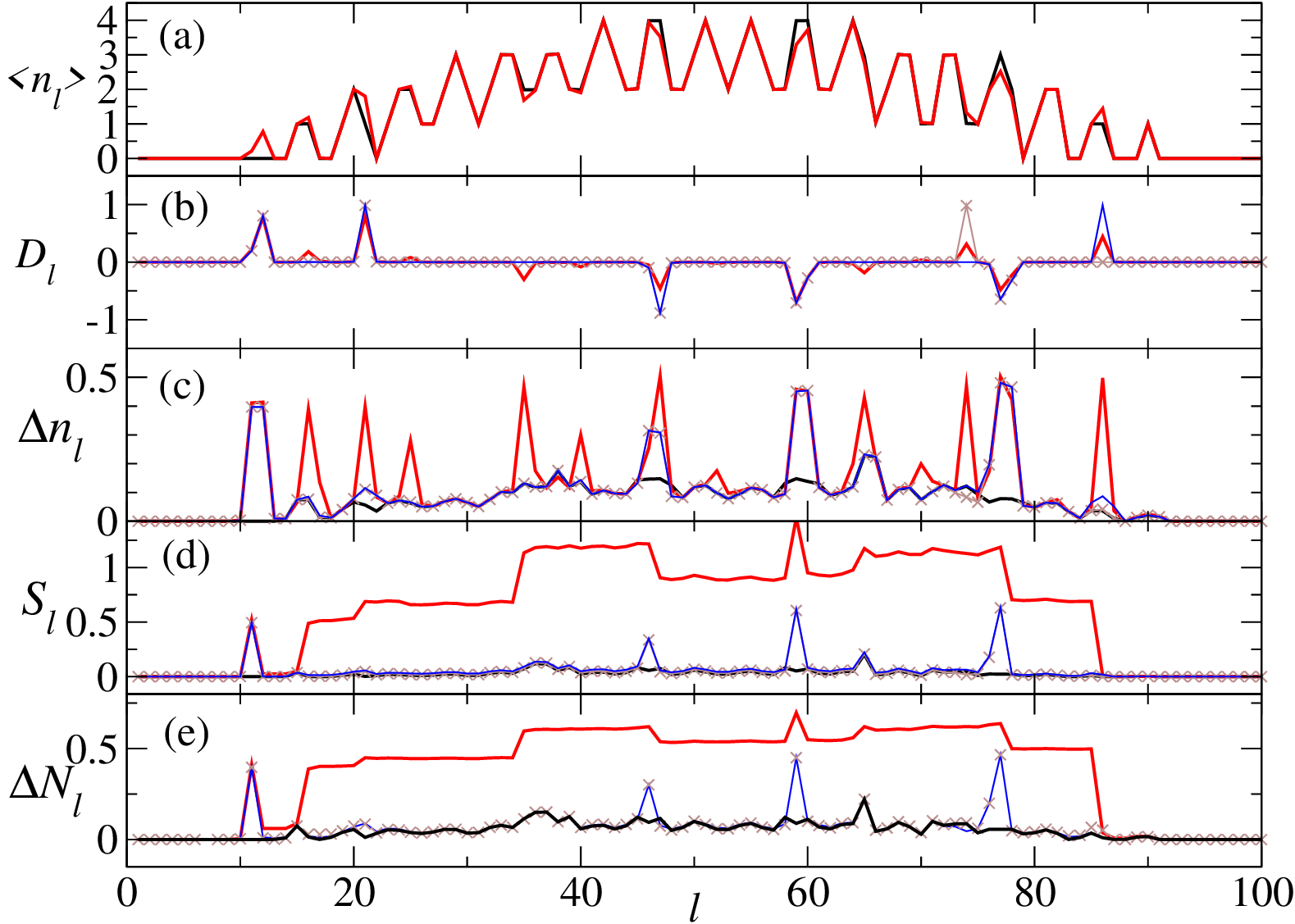}
\end{center}
\caption{(color online)
Properties of states of bosons trapped in
a deep optical lattice ($s$=14) in the presence of strong disorder $s_2$=2.1875.
Here
$r_0$=0.1075 and $\phi$=0.0304 in Eq.(\ref{trap})).
Brown (blue) lines correspond to excitations X and Y 
in Fig.~\ref{fig:autocorr_spectra}(c).
The entanglement entropy is much larger for the dynamically created wavepacket than for stationary states, 
and $\Delta n_l$ has many more peaks, indicating a~significant melting of the Bose glass.
\label{fig:full_s25}}
\end{figure}

For strong disorder, $s_2=2.1875$, no MI state exists and the ground state of the system forms
a BG. The exemplary excitations are shown in Fig.~\ref{fig:full_s25}. 
The occupations of various sites strongly fluctuate, 
low lying excitations seem quite similar to the ground state. 
Excitations are {\it local} in character modifying the occupation number and its variance in selected 
(not always close) sites only.
The entanglement entropy is small, except in small SF pockets. 
Again, the wavepacket has large entanglement entropy and $\Delta n_l,$ with numerous peaks
indicating melted regions. 

\section{Conclusions}

In conclusion, we have built an extension of the TEBD algorithm 
capable of extracting, from a non stationary state (created dynamically), the 
eigenstates dynamically populated, and
the nature of the
``defects'' created.  In the specific case of the BH model realized in a
ultracold atomic gas, a quasi-adiabatic quench populates excited states that
are essentially similar to the ground state.
In contrast, the dynamical wavepacket has a~markedly different character, with 
a~significant entanglement across the whole sample. 
By modulation spectroscopy, higher excited states may also be selectively reached
and analyzed.

It is clear that our method will not provide an access to all excited states of
a given many body system. By construction, only those excited states which are well represented 
by a MPS within a given assumed size of the internal Hilbert space $\chi$
may be obtained. 
On the other hand, the TEBD algorithm seems to create a ``self-purification'' 
mechanism which enables to reach integration times much longer than originally
expected \cite{vidal03,daley}. Naturally the result of this integration is not a good 
approximation of the true dynamics but contains information about the well represented excited states. 
Thus our method seems to have larger range of applicability than expected on the basis of 
standard limitations of the TEBD algorithm.
 
 In that context, wavepackets with 
 low excess energy w.r.t. the ground state created via quasi-adiabatic 
quench simulating the Florence experiment with disorder \cite{fallani07} behave exceptionally well. Higher excitations reached in modulation spectroscopy were more computer intensive and revealed
that the method cannot be used as a kind of ``black-box'' approach but has to be tuned to 
individual cases. While this is certainly some drawback, the method potential capabilities 
(as exemplified with the highly excited state shown in fig.~\ref{fig:kollmo}d) seem quite encouraging.

\section{Acknowledgements}

We are grateful to M. Fleischhauer and his colleagues for critical remarks on an early
 version of this paper. 
Support within Polish Government scientific funds for 2009-2012
as a~research project is acknowledged. M{\L} acknowledges
support from Jagiellonian University International Ph.D
Studies in Physics of Complex Systems (Agreement No.
MPD/2009/6) provided by Foundation for
 Polish Science and cofinanced by the European  Regional Development Fund. 
Computer simulations were performed at supercomputer Deszno, 
purchased thanks to the financial support of the European Regional Development Fund 
in the framework of the Polish Innovation Economy Operational Program, 
contract no. POIG.02.01.00-12-023/08 (JZ),
ACK Cyfronet AGH as a~part of the POIG PL-Grid project (M\L) 
and at ICM UW under Grant No. G29-10 (JZ and M\L). 

\section{Appendix:  implementation of MPS-addition}

The addition of two MPS, an elementary step in the FT, produces $\lambda$ vectors with increased size.
Thus it seems that the $\chi$ value required for representing successive sums 
should increase fast, ruining the efficiency
of the MPS representation. This is not the case: 
the $\lambda^{[l]}_{\alpha_l=1,2\ldots}$ for the partial
sums decrease relatively fast so that the smallest components 
can be safely discarded~\cite{verstraete}, preventing the cutoff to
blow up, a~key point for the success of the method. 
Altogether, several thousands states can be summed, at the price of roughly
doubling the $\chi$ value to preserve
the accuracy of the MPS representation: when the series is dynamically created, as in eq.~(\ref{eq:phiw}), 
the result is close to an eigenstate, i.e. simpler than the
summands which are wavepackets, and the vectors added together are closely related. 

To be able to compute the sum of several thousands of MPS-vectors two issues have to be resolved: 
restoring the canonical form of a MPS vector and performing truncation in $\chi.$ 
The former 
has been described in \cite{shividal}. The truncation from 
$\chi=\chi_2$ to $\chi=\chi_1$ can be done in two nonequivalent ways: ''blunt''
and ''canonical''. The canonical truncation is done in parallel with restoring the canonical form.  
The algorithm described in
\cite{shividal} processes sites sequentially restoring orthogonality relations at each of them
one at a time. The ''left'' and ''right''  Schmidt vectors for site $k$ are:

\begin{equation}
|\psi^{1\ldots k}_{\alpha_k}\rangle = \!\!\!\!\!\!\sum\limits_{\genfrac{}{}{0pt}{}{\alpha_1,\ldots,\alpha_k}{i_1,\ldots,i_k}}\!\!\!\!\!
\Gamma_{1\alpha_1}^{[1]
,i_1}\lambda^{[1]}_{\alpha_1}\Gamma^{[2],i_2}_{\alpha_1\alpha_2}\ldots\Gamma^{[k],i_k}_{\alpha_{k-1}\alpha_k} |i_1,\dots,i_k\rangle
\label{eqn:MPS2}
\end{equation}
\begin{equation}
|\psi^{k+1\ldots M}_{\alpha_k}\rangle = \!\!\!\!\!\!\sum\limits_{\genfrac{}{}{0pt}{}{\alpha_{k+1},	\ldots,\alpha_M}{i_{k+1},\ldots,i_M}}\!\!\!\!\!
\Gamma_{\alpha_k\alpha_{k+1}}^{[k+1]
,i_{k+1}}\lambda^{[k+1]}_{\alpha_{k+1}}\ldots\Gamma^{[M],i_M}_{\alpha_{M-1}\alpha_1} |i_{k+1},\dots,i_M\rangle
\label{eqn:MPS3}
\end{equation}
\begin{equation}
|\psi\rangle = \sum\limits_{\alpha_k}\lambda_{\alpha_k}^{[k]} |\psi^{1\ldots k}_{\alpha_k}\rangle\otimes |\psi_{\alpha_k}^{k+1,\ldots,M}\rangle
\end{equation}
For a MPS vector in its canonical form, these vectors are orthonormal eigenvectors of 
the reduced density matrices
$\rho^{[1\ldots,k]},$ $\rho^{[k+1\ldots,M]}.$ The algorithm in \cite{shividal} restores orthonormality 
of left and right
Schmidt vectors at site $k$ assuming it has already been applied for sites $1,\ldots,k-1.$ 
After the orthogonality has been restored
at site $k,$ and the $\lambda^{[k]}$ sorted by decreasing magnitude, 
we zero all $\lambda^{[k]}_{\alpha_k}$ for
$\chi_1 <\alpha_k \leq \chi_2.$ The truncation error is estimated by $\sum_{\alpha_k=\chi_1+1}^{\chi_2}
(\lambda^{[k]}_{\alpha_k})^2$ \cite{verstraete}. After all sites have been processed, 
at each site left Schmidt vectors are
orthonormal eigenvectors of the appropriate reduced density matrix.  
Orthonormality of right Schmidt vectors at site $k$ is lost
when truncation of $\lambda^{[k+1]}$ takes place. If the algorithm were applied again 
(without truncation), both left and
right Schmidt vectors would be canonical for any site. 
We do not perform this second part since we do not use orthonormality of
right Schmidt vectors for the addition procedure.

The blunt way to truncate the MPS state is to first use the previous algorithm with no parallel truncation at all, only then
perform truncation of the vectors $\lambda^{[k]}$ at all sites. 
This method is mathematically flawed as, starting from the
second site truncation, the canonical form is lost.  
However, when the $\chi$ value used for FT evaluation is larger than $\chi$ for the
temporal evolution, both methods agree well.

In that respect, wavepackets with low excess energy w.r.t. the ground state created 
via quasi-adiabatic quench simulating the Florence experiment with disorder \cite{fallani07} 
behaved exceptionally well. 
The converged results could be obtained with $\chi=80$ as tested and compared with $\chi=120$ or $\chi=150$.
  
In contrast, extraction of selected eigenstates from wavepackets created by
harmonic modulation is more demanding. 
Too strong modulation creates wavepackets containing apparently strongly entangled contributions. 
In such a case, a long time propagation of such a wavepacket is not possible 
even for large $\chi$ values. 
Lowering the amplitude of the modulation and adjusting its frequency is a key ingredient 
for successful extraction of a given excited state. 
The results presented in this paper have been obtained for $\chi=120$ 
again testing the convergence by comparison with $\chi=150$.

The convergence may be also tested by monitoring the growth of the smallest $\lambda^{[l]}_{\chi}$ values.

%%%%%%%%%%%%%%%%%%%%%%%%%%%%%%%%%%%%%%%%%%%%%%%%%%

%%%%%%%%%%%%%%%%%%%%%%%%%%%%%%%%%%%%%%%%%%%%%%%%%%

\end{document}